\documentclass[aps,prd,eqsecnum,amsmath,amssymb,longbibliography,noeprint,10pt,twocolumn,showkeys]{revtex4-2}

\usepackage{graphicx}
\usepackage{dcolumn}
\usepackage{bm}
\usepackage{orcidlink}  
\usepackage{subcaption}
\usepackage[table,xcdraw,dvipsnames]{xcolor}
\usepackage{ragged2e}
\usepackage{soul}
\usepackage{url}
\usepackage{hyperref}

\newcommand{\apjl}{Astrophysical Journal Letters}

\newcommand{\aap}{Astronomy \& Astrophysics}

\definecolor{blue-violet}{rgb}{0.7, 0.2, 0.8}

\sethlcolor{yellow}  
 
\begin{document}

\preprint{AIP/123-QED}

\title{Finding Supermassive Black Hole Binary Mergers in Pulsar Timing Array Data}

\author{Sharon Mary Tomson \orcidlink{0000-0001-7603-1637}}
 \email{sharon.mary.tomson@aei.mpg.de}
\author{Boris Goncharov \orcidlink{0000-0003-3189-5807}}%
\author{Rutger van Haasteren \orcidlink{0000-0002-6428-2620}}
\affiliation{ 
Max Planck Institute for Gravitational Physics (Albert Einstein Institute), 30167 Hannover, Germany\\
Leibniz Universität Hannover, 30167 Hannover, Germany
}%

\date{\today}

\begin{abstract}

Galaxy observations suggest that mergers of supermassive black hole binaries (SMBHBs) are rare events on cosmological scales, with characteristic rates of order one per decade across the observable Universe. In this work, we present a framework to search for merging SMBHBs in pulsar timing array (PTA) data using a physically complete waveform model that includes the late inspiral, merger, ringdown, and gravitational-wave memory. This approach enables a unified treatment of continuous pre-merger emission and the non-oscillatory memory signal associated with coalescence. Using simulated PTA datasets, we demonstrate a proof of principle for parameter estimation of merging SMBHBs, focusing on representative systems with chirp masses of $10^{8}$ and $10^{10}~M_\odot$ at distances of 3 Mpc and 100~Mpc respectively. These signals are distinguishable for a PTA consisting of 25 pulsars observed over 13 years with 100 ns timing precision. For these sufficiently strong signals, we recover the simulated binaries with log Bayes factors exceeding $10$, and show that chirp mass and luminosity distance can be jointly constrained with uncertainties governed by the characteristic mass–distance degeneracy of the signal.
The sky position uncertainties are of the order of a few degrees, which could potentially enable electromagnetic follow-up and multi-messenger observations of SMBHB mergers. We also show that the measurement uncertainties on the parameters of simulated merging binaries depend weakly on the presence of the gravitational wave background with Hellings-Downs correlations in our simulated data.
We further demonstrate that commonly used memory burst approximations lead to biased strain amplitudes and inferred source parameters when compared to the full SMBHB waveform, even when optimally tuned. These results establish a pathway toward searching for SMBHB mergers and late-stage inspirals with PTAs using physically complete waveform models, complementing existing searches for continuous gravitational waves.

\end{abstract}

 
\maketitle

\section{\label{sec:intro}Introduction}

Pulsar Timing Arrays (PTAs) have the capability to detect gravitational waves (GWs) in the nanohertz frequency regime by precisely monitoring variations in the times of arrival (TOAs) of radio pulses emitted by millisecond pulsars (MSPs)~\citep[][]{FosterBacker1990}. These subtle fluctuations in TOAs are caused by GWs propagating through spacetime, perturbing the metric and inducing correlated timing deviations in widely separated pulsars. By analyzing TOAs, PTAs provide a unique means of probing low-frequency gravitational radiation.

Currently, several PTA collaborations are active: the North American Nanohertz Observatory for Gravitational Waves~\citep[NANOGrav,][]{McLaughlin2013}, the Parkes Pulsar Timing Array~\citep[PPTA,][]{ManchesterHobbs2013}, the European Pulsar Timing Array~\citep[EPTA,][]{DesvignesCaballero2016}, the Indian Pulsar Timing array~\citep[InPTA,][]{TarafdarNobleson2022}, the Chinese PTA~\citep[CPTA,][]{Lee2016} and the MeerKAT PTA ~\citep[MPTA][]{MilesShannon2023}. 
The International Pulsar Timing Array~\citep[IPTA][]{VerbiestLentati2016} is a consortium of PTAs.

It is expected that the stochastic gravitational wave background (GWB) will become the first gravitational wave signal observed by PTAs.
Multiple lines of evidence for the background have been recently reported~\citep{NG_12_GWB,PPTA_DR2_GWB,GoncharovThrane2022,EPTA_CRN,IPTA_DR2_GWB,NG_15_GWB,EPTA_DR2_GWB,PPTA_DR3_GWB,MT_DR1_GWB,CPTA_DR1_GWB}.
Such a background is expected to originate from a population of inspiraling supermassive black hole binaries (SMBHBs), although the early-universe origins are also not ruled out ~\citep{NG_15_NEWPHYS,EPTA_SGWB}. 

In addition to the background, PTAs also search for deterministic signals which could be brighter and individually resolvable, such as the continuous gravitational waves (CGWs) from individual SMBHBs~\citep{EllisJenet2012}, burst events from cosmic string cusps \cite{DamourVilenkin2000, XiaWang2025}, SMBHB mergers or hyperbolic binaries \citep{Turner1977}, and the gravitational wave memory \citep{Favata2010,vanHaasterenLevin2010,DandapatSusobhanan2024}. Recently, PTAs have begun targeted CW searches of active galactic nuclei (AGN) candidates selected from electromagnetic (EM) surveys, using EM-informed priors on sky location, distance/redshift, and CW frequency~\citep{AgarwalAgazie2025}. Complementing these targeted searches, morphology-agnostic methods for generic burst signals have recently been developed for PTA data sets, enabling efficient, model-independent burst detection in the nanohertz band~\citep{TaylorBurnette2025}.

In this context, mergers of SMBHBs are of particular interest because they produce a component of the gravitational wave signal referred to as memory. 
The most prominent memory component, the \textit{displacement memory}, manifests as the accumulating displacement of the freely-falling test particles as a gravitational wave passes by. Theoretical work over the past few decades has classified memory, including displacement memory, into two main types: ordinary and null ~\cite{Favata2010}.

Ordinary memory -- also called linear memory -- occurs in systems where matter is ejected from the system: hyperbolic encounters of compact objects~\cite{Turner1977, DandapatEbersold2023}, asymmetric neutrino emission from supernovae~\cite{Epstein1978, MadisonChernoff2017}, or relativistic jets from gamma-ray bursts~\cite{SagoIoka2004}.
Ordinary memory was first discovered in the 1970s by considering hyperbolic encounters of compact objects~\cite{Zel'dovichPolnarev1974, BraginskyThorne1987}. 
Search for linear memory from SMBHBs on hyperbolic trajectories has been performed in the NANOGrav 12.5-year dataset, finding no compelling evidence and placing limits on the existence of such systems~\citep{DandapatSusobhanan2024}. 
Ordinary memory is negligible for SMBHB mergers. 

Null memory, previously mostly referred to as nonlinear or Christodoulou~\citep{Christodoulou1991} memory, is due to the contribution of gravitational radiation itself to the evolution of the system's changing quadrupole and higher mass moments~\cite{Favata2009}. 
This implies that almost all GW sources inherently produce some level of memory. 
For quasi-circular compact binary mergers, which are among the most significant and well-studied sources, the null memory contributes substantially to the time-domain waveform. 
In the post-Newtonian (PN) framework, the memory enters at the Newtonian (leading) order. This arises because the memory is a hereditary effect—it depends not just on the current configuration of the source, but on its entire past history. Furthermore, its non-oscillatory behavior makes it stand out in the waveform, providing a clear signature distinct from the oscillatory inspiral and merger phases.
Because of these properties, null memory is not only a powerful probe of GW sources but also serves as a unique manifestation of general relativity’s nonlinear structure. This effect, first derived in full by Christodoulou~\cite{Christodoulou1991}, remains one of the most profound consequences of gravitational wave theory, highlighting how spacetime curvature can be permanently altered by the radiation it carries. 

Current PTA searches for null memory largely rely on a signal model in which the memory-induced metric change is treated as an instantaneous, step-like shift~\citep{vanHaasterenLevin2010}. 
This model, from now on referred to as a \textit{memory burst}, models the GW memory as a ramp in timing residuals that begins at the time of the burst and persists thereafter. 
The justification for this approximation arises from the fact that the memory grows over a timescale of a few days to weeks for SMBHBs with total mass $M\sim 10^{8}~M_{\odot} - 10^{10}~M_{\odot}$. Compared to the decade-long observing span and typical bi-weekly cadence of PTA experiments, this growth appears instantaneous, and the detailed time structure of the signal becomes unresolved. This approximation has been used in several pioneering studies \citep{vanHaasterenLevin2010,Seto2009,PshirkovBaskaran2010}. 
The detectability of such bursts using the memory burst model with 20 simulated pulsars timed at 100 ns precision over 10 years is investigated in \citet{vanHaasterenLevin2010}.
The authors concluded that an SMBHB merger with  $M = 10^{8}~M_{\odot}$ at $z=0.1$ could be detected with $2\sigma$ confidence, whereas $M = 10^{10}~M_{\odot}$ would be detectable throughout the Universe. 
Searches for gravitational wave memory that have been carried out on real PTA datasets include the frequentist approach using the memory burst model on the first PPTA dataset  \citep{WangHobbs2015} and the Bayesian approach by NANOGrav on their 11-,12.5-, and 15-year datasets~\citep{NG_11_MEM, NG_12_MEM, NG_15_MEM}. 
Although no definitive detections have been reported, the 15-year analysis showed a mild preference (Bayes factor $\approx$ 3) for a model including a memory burst.

While searches for memory bursts apply to multiple physical sources, there are compelling reasons to move beyond this approximation. 
First, the burst template is derived under the objective to search for memory alone. 
This overlooks a key point: null memory arises as a consequence of the changing quadrupole moment brought out by the oscillatory gravitational waveform.  
It is not a standalone phenomenon, but rather a byproduct of the same quadrupolar dynamics that generate the inspiral, merger, and ringdown phases. 
These oscillatory components, particularly for nearby or massive binaries, may themselves imprint detectable signatures in PTA timing residuals. 
Second, the burst model’s extreme simplicity makes it vulnerable to unmodeled noise. 

In this work, we introduce the first complete signal model for SMBHB mergers that uses a full inspiral–merger–ringdown (IMR)  numerical relativity surrogate which also includes null memory. This goes beyond the post-Newtonian, inspiral-only waveforms used in prior PTA continuous-wave searches. Using a high-fidelity NR waveform~\cite{YooMitman2023}, we accurately model the full time-domain strain, including memory.

By simulating this signal in PTA datasets, we demonstrate how incorporating realistic waveform features enhances detection prospects and improves parameter estimation. Relying solely on a generic burst model discards the rich astrophysical information encoded in the full waveform. In contrast, a physically motivated model enables direct inference of key source properties such as chirp mass, mass ratio, luminosity distance, and inclination from the data.

The rest of the paper is organized as follows. 
In Section~\ref{sec:method}, we outline the methodology of our search, introduce the signal model and quantities we can infer. 
In Section~\ref{sec:results}, we present our results on simulated datasets, sensitivity of PTA to SMBHB mergers and limitations of memory burst model. 
In Section~\ref{sec:discussion}, we discuss improvements to the memory burst model, the new window opened onto SMBHBs for PTAs, future prospects of the SMBHB merger model and its potential for enabling EM follow-ups. We summarize the conclusions for the complete signal model in Section~\ref{sec:conclusion}.

\section{\label{sec:method}Methodology}

\subsection{\label{sec:model}Memory Signal from a Binary Merger}

In this subsection, we provide background information about the calculation of gravitational wave memory strain.
This information is useful for understanding the relation of the gravitational wave memory signal to properties of SMBHB, as well as to motivate our methodology. 

Expressing the gravitational waveform time series $h_+ + i h_\times$ as the sum of the spherical harmonics with spin weight of $-2$, the leading contribution to null displacement memory from quasi-circular binaries arises from the amplitude of the dominant $h_{22}$ mode of the oscillatory component of the signal~\cite{Favata2010}. 
Assuming a non-precessing binary with component spins aligned with the orbital angular momentum, the memory can be computed analytically by integrating the square of the time derivative of this mode as 
\begin{equation}
\label{eq:memhplus}
h_{+}^{\rm (mem)} \approx \frac{R}{192\pi} s_{\iota}^2 (17+c_{\iota}^2) \int_{-\infty}^{T_R} |\dot{h}_{22}|^2 \mathrm{d} t,
\end{equation}
where $R$ is the distance from the source to the observer, $\iota$ is the angle between the line of sight and the orbital angular momentum of the binary system,  $c_{\iota} \equiv \cos\iota$, $s_{\iota}\equiv \sin\iota$, and $T_R$ is the retarded time.

For compact binaries including SMBHBs, Equation~\ref{eq:memhplus} evolves into a minimal waveform model (MWM) expression as explained by Favata~\citep{Favata2009} as 
\begin{multline}
\label{eq:hmemMWM}
{h}^{\rm (mem)}_{\rm MWM}(T) = \frac{8\pi M}{r(T)} \Theta(-T) +  \Theta(T) \Bigg\{ \frac{8\pi M}{r_m}   + \frac{1}{\eta M} \\ \!\!\!\! \times \!\!\!\!  \sum_{n,n'=0}^{n_{\rm max}} \!\!\!  \frac{\sigma^{\,}_{22n}\sigma_{22n'}^{\ast} A^{\,}_{22n} A_{22n'}^{\ast}}{\sigma^{\,}_{22n} + \sigma_{22n'}^{\ast}}  \left[ 1 - e^{-(\sigma^{\,}_{22n} + \sigma_{22n'}^{\ast})T} \right] \!\! \Bigg\},
\end{multline}
where $\Theta(t)$ is the Heaviside function, $M$ the total mass, $r(T)$ the binary separation at time T, $\eta$ the symmetric mass ratio, $A_{22n}$ the amplitudes of the quasinormal modes (QNMs) of the dominant $\textit{l} = 2$ , $m=2$ GW mode during ringdown , $\sigma_{22n}$ being the complex QNM frequencies and $r_m \equiv r(T=0)$ the orbital separation at the end of the inspiral, where the inspiral and ringdown descriptions are matched (refer \cite{Favata2009} for more details). This resulting expression captures the gradual accumulation of memory during the inspiral and merger, analytically. 

Favata also developed a more relativistic Effective One Body (EOB) waveform model calibrated to numerical relativity (NR) simulations shown in Fig.~1 of~\cite{Favata2009}. Although the EOB framework offers improvements over the MWM, its predictions still deviate from full NR-based calculations. For instance,  the final memory amplitude from a hybrid PN/NR waveform was found to differ by 27\% compared to the EOB estimate \cite{Favata2009}.

PTA searches have often relied on a much simplified representation of memory -- a step function approximation in the metric perturbation. 
In the limit where the null memory saturates rapidly compared to the sampling interval, the detailed expression in equation \eqref{eq:hmemMWM} is approximated as a sudden change in the metric traveling through space \cite{vanHaasterenLevin2010},
\begin{equation}
    h_{+}(\vec{r}, t)=h_0\times \Theta\left[(t-t_0)-\vec{n}\cdot \vec{r}\right],
    \label{eq:dcwave}
\end{equation}
where $h_0$ is the amplitude of the step, $t_0$ is the time at which the step passes the Earth, $\vec{r}$ is the position vector of the observer, and $\vec{n}$ is the unit vector in the direction of wave propagation. 

Even though the above approximations are well-motivated and sound, they fall short in capturing the full physical accuracy of the SMBHB signal. 
As noted by~\citep{Favata2009}, the step function and the EOB waveform model tend to overestimate the final memory saturation value compared to NR waveforms. Thus, to be able to confidently detect SMBHB mergers and perform a robust parameter estimation, full NR-informed waveforms that capture both the oscillatory and non-oscillatory contributions are necessary. 

To address the above, we adopt the state-of-the-art numerical relativity surrogate waveform model \texttt{NRHybSur3dq8\_CCE} of~\citet{YooMitman2023}.
It is built using the Cauchy-characteristic evolution (CCE), a method that evolves the gravitational field fully to future null infinity, thereby avoiding gauge ambiguities and capturing gravitational memory effects that are lost in more common extrapolation-based waveform extraction \citep{BishopGomez1997, MoxonScheel2023, MitmanMoxon2020}. 
The model includes both the oscillatory (inspiral-merger-ringdown, IMR) and null memory components of the strain, thereby providing a complete description of the expected waveform from SMBHBs.

In \citet{YooMitman2023}, the CCE-extracted numerical-relativity (NR) waveforms are first placed into the same asymptotic frame used by post-Newtonian (PN) theory, known as the Bondi–Metzner–Sachs (BMS) frame. This step removes residual gauge differences between the NR and PN descriptions. The early inspiral is modeled using analytic waveforms based on post-Newtonian (PN) expansions informed by Effective One-Body (EOB) dynamics. This single analytic inspiral model is then smoothly hybridized with the NR waveform, producing a continuous signal spanning inspiral, merger, and ringdown.

The model is parametrized in terms of chirp mass $\mathcal{M}$, mass ratio $q$, effective and antisymmetric spin $\chi_\text{eff}$, luminosity distance $D_\text{L}$, inclination, polarization and sky location of the source. 
The \texttt{NRHybSur3dq8\_CCE} model accurately reproduces the CCE-extracted NR waveforms used for training, as demonstrated in Ref.~\cite{YooMitman2023}.
To construct the gravitational wave (GW) signal model as observed by a pulsar timing array (PTA), we compute the two strain polarization components, $h_{+}(t)$ and $h_{\times}(t)$ using the \texttt{NRHybSur3dq8\_CCE} waveform. In this analysis we include the oscillatory modes  $ (l,m) = {(2\pm2),(2\pm1),(3\pm3)}$ which describe the IMR part of the signal, together with the non-oscillatory $(2,0)$ mode, which carries the gravitational-wave memory contribution (present only in the $h_{+}(t)$ polarization).

\subsection{Projection in Pulsar Timing Arrays}

\label{sec:method:projection}

The response of a PTA to the shift in pulsar rotational frequency $\nu$ due to gravitational wave strain $h$ with $+$ and $\times$ polarizations is given by \citet{vanHaasterenLevin2010}:
\begin{equation}
\label{eq:freqshift}
{\frac{\delta \nu(t)} {\nu}}= F(\hat{\Omega}, \hat{p}, \psi)\times \left[ h_{+,\times}(t) - h_{+,\times}(t- \vec{n} \cdot \vec{r})\right],
\end{equation}
where $F(\hat{\Omega}, \hat{p}, \psi)$ is the geometric projection factor which depends on $\hat{p}$, the unit vector pointing from the Earth to the pulsar,  $\hat{\Omega}$, the unit vector pointing from the GW source to the Solar System Barycenter (SSB), and $\psi$ the gravitational wave polarization angle. The angle between the source and pulsar directions is defined as $\cos{\mu} = - \hat{\Omega} \cdot \hat{p}$.
The first strain term in Equation~\eqref{eq:freqshift} is the ``Earth term'' and the second is the ``pulsar term''. 
The Earth term is a coherent signal that appears simultaneously across all pulsars. 
The PTA response corresponding to the rotational frequency change is obtained by integrating Equation~\eqref{eq:freqshift}:
\begin{equation}
 \label{eq:residuals}
    {\delta t (t) = \int_{t_{0}}^{t} {\frac{\delta \nu(t')} {\nu}} dt'}.
  \end{equation}

It is also convenient to rewrite the Earth term of dimensionless gravitational wave strain $h(t)$ seen in Equation~\eqref{eq:freqshift} in the matrix form:
\begin{align}
h(t)&=\begin{bmatrix}F_{+} & F_{\times}\end{bmatrix}\begin{bmatrix}\cos2\psi & -\sin2\psi\\
\sin2\psi & \cos2\psi
\end{bmatrix}\begin{bmatrix}h_{+}(t)\\
h_{\times}(t)
\end{bmatrix},\
\label{eq:h_matrix}
\end{align}
where $\psi$ is the polarization angle of the source and $h_{+,\times}$ denote the two independent polarization modes of the strain in the transverse-traceless gauge. 

The functions $F_{+,\times}$ are the antenna pattern functions defined as
\begin{align}
F_{+}(\hat{\Omega})&=\frac{1}{2}\frac{(\hat{m}\cdot\hat{p})^{2}-(\hat{n}\cdot\hat{p})^{2}}{1+\hat{\Omega}\cdot\hat{p}}\\
F_{\times}(\hat{\Omega})&=\frac{(\hat{m}\cdot\hat{p})(\hat{n}\cdot\hat{p})}{1+\hat{\Omega}\cdot\hat{p}},
\label{eq:antennapattern}
\end{align}
$\hat{m}$ and $\hat{n}$ are orthonormal basis vectors spanning the plane transverse to the GW propagation direction $\hat{\Omega}$, defined as~\citep{EllisSiemens2012}
\begin{align}
\hat{m} &= (\sin\phi, -\cos\phi, 0)
\end{align}
\begin{align}
\hat{n} &= (-\cos\theta\cos\phi, -\cos\theta\sin\phi, \sin\theta)
\end{align}
with $(\theta,\phi)$ being the sky location of the GW source.

The first two matrices in Equation~\eqref{eq:h_matrix} are the expansion of $F(\hat{\Omega}, \hat{p}, \psi)$ in Equation~\eqref{eq:freqshift}.
The response on a pulsar $\delta t (t)$ is then expressed as 
\begin{align}
    \delta t (t) &= \begin{bmatrix}F_{+} & F_{\times}\end{bmatrix}\begin{bmatrix}\cos2\psi & -\sin2\psi\\
\sin2\psi & \cos2\psi
\end{bmatrix}\begin{bmatrix}\delta t_{+}(t)\\
\delta t_{\times}(t)
\end{bmatrix}\,.
\label{eq:r_matrix}
\end{align}
where functions $\delta t_+ (t)$ and $\delta t_\times (t)$ represent the time-integrated gravitational wave strain as in Equation~\eqref{eq:residuals}:
\begin{align}
\label{eq:st}
  \delta t_{+,\times}(t) & =   \int_{t_0}^t \, h_{+\times} (t')\, dt'.
\end{align}
This formalism applies to any gravitational wave model, whether it describes a continuous gravitational wave or a memory burst. In this work we model only the Earth term, which is coherent across pulsars and provides the correlated signature used for detection and sky localization. The pulsar term depends on the uncertain pulsar distance and for gravitational wave memory it is generally displaced by $ L ( 1 + \hat{\Omega} \cdot \hat{p}) $ placing the associated step outside a typical decade-long observing span for kpc-distance pulsars. When the pulsar term occurs within the data span, it yields at most a single step (jump) in the GW strain, localized to an individual pulsar, which can be difficult to distinguish from intrinsic pulsar noise processes such as pulsar glitches~\citep{McKeeJanssen2016}.

\begin{figure*}
  \centering
  \begin{subfigure}{0.32\textwidth}
    \includegraphics[width=\textwidth]{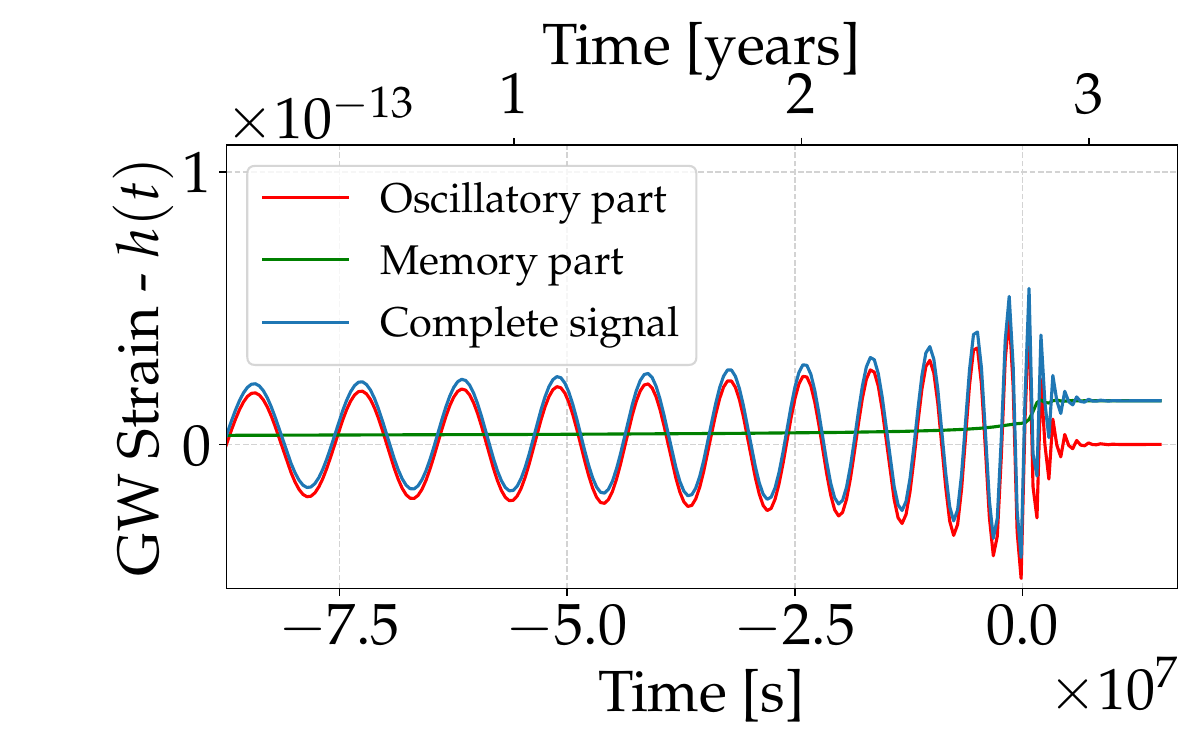}
    \caption{}
    \label{fig:NR_waveforms:ht}
  \end{subfigure}
  \hfill
  \begin{subfigure}{0.32\textwidth}
    \includegraphics[width=\textwidth]{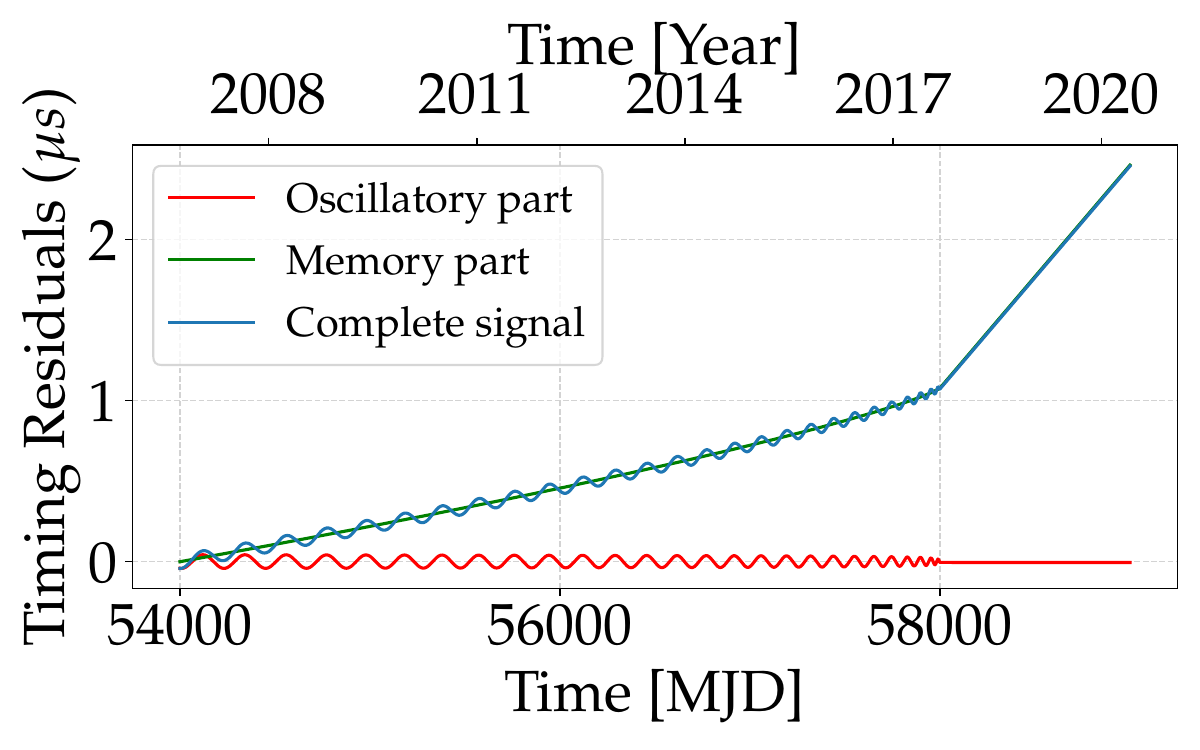}
    \caption{}
    \label{fig:NR_waveforms:res}
  \end{subfigure}
  \hfill
  \begin{subfigure}{0.32\textwidth}
    \includegraphics[width=\textwidth]{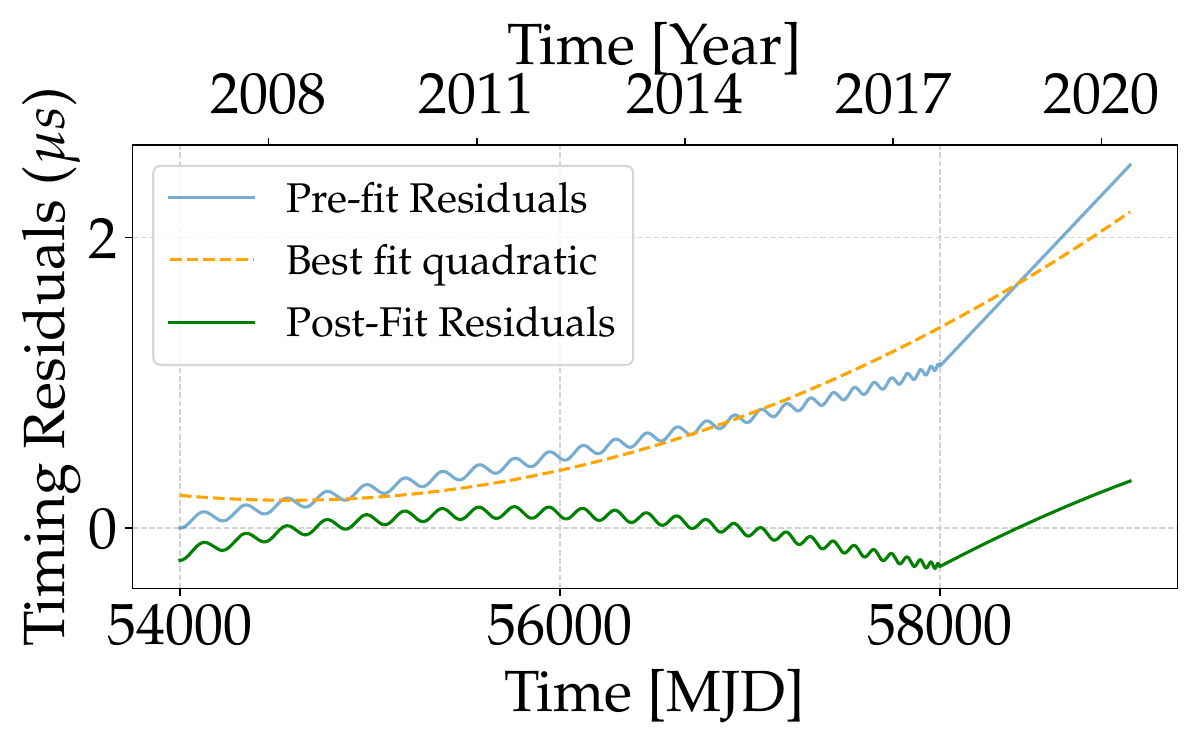}
    \caption{}
    \label{fig:NR_waveforms:quad}
  \end{subfigure}
  \caption{\justifying Gravitational wave strain and timing residuals from a merger of a non-spinning supermassive black hole binary (SMBHB) with parameters
  $\mathcal{M}_c = 10^{10} M_{\odot}$, $q=1$, and $D_\text{L} = 1000$ Mpc.
  (a) GW strain waveform $h(t)$ for three different combinations of radiative modes:
  $(l,m) = (2\pm2),(2\pm1),(3\pm3)$ without memory (red), the null memory-only $(2,0)$ mode (green), and the complete signal including both oscillatory and memory contributions (blue).
  (b) The corresponding response induced by the GW source located at $(\text{ra},\text{dec}) = (0^{\circ},0^{\circ})$ on a pulsar at
  $(\text{ra},\text{dec}) = (258.4564^{\circ},7.7937^{\circ})$. Colours match those in (a) and the merger occurs at MJD 58000.
  (c) Post-fit residuals after subtraction of the best-fit pulsar spin frequency and its derivatives. The full signal is shown in blue, the linear and quadratic spin-down model in orange, and the resulting post-fit residuals in green.}
  \label{fig:NR_waveforms}
\end{figure*}

After computing the GW strain polarizations, $h_+(t)$ and $h_\times(t)$ using the surrogate waveform model, the timing residuals are then calculated using the formalism in Equation~\eqref{eq:r_matrix}.
It is demonstrated in Figure~\ref{fig:NR_waveforms}.
Figure~\ref{fig:NR_waveforms:ht} shows the resulting strain waveform for various mode combinations: the dominant oscillatory modes without memory (in which the strain $h(t)$ decays to zero after the merger), the memory only $(2,0)$ mode, and the full waveform including both the IMR and memory contributions. 
The corresponding response in the timing residuals for an arbitrary pulsar located at a right ascension of $258.4564^{\circ}$ and a declination of $7.7937^{\circ}$ is shown in Figure~\ref{fig:NR_waveforms:res}. The SMBHB source is at a right ascension of $0^{\circ}$ and a declination of $0^{\circ}$. The observation spans from MJD 54000 to 59000, covering approximately 13 years, with the merger occurring at MJD 58000. Figure~\ref{fig:NR_waveforms:quad} shows the post-fit residuals obtained after the subtraction of the best-fit pulsar spin frequency and its first derivative.

To construct the signal model for use in PTA analysis, we generate $h(t)$ over an extended duration of 30 years, which is larger than the observation spans of the analyzed data. 
This ensures that the full signal, including the pre-merger oscillations and memory buildup,
is captured regardless of when the binary coalescence occurs relative to the observation window. 
Such extrapolation is particularly important when searching for SMBHBs whose merger might occur outside the current PTA timespan. The waveform is computed starting from a frequency determined by the total binary mass and chosen duration (30 years before merger). 
The resulting waveform is interpolated to match the sampling time of the PTA data. 
To get the timing residual, we integrate the strain using a cumulative trapezoidal integration of $h(t)$. 
Post-merger, the memory component is extended as a constant offset based on the final saturation value of  $h(t)$, similar to the memory burst model. 
This allows the signal to smoothly continue after the merger epoch.  
The implementation of the waveform generation and timing residual projection routines used in this work is publicly available in~\cite{tomson_github_smbhb_memory}.

\subsection{Data analysis} 
\label{sec:method:analysis}

Throughout this study, we simulate pulse arrival time observations with an uncertainty on pulse arrival times of $\sigma_\text{TOA}=100$~ns. 
Such low uncertainties are already achieved, and in some cases even surpassed, for several well-timed millisecond pulsars \citep{MilesShannon2025}. 
We simulate pulse arrival time observations at 1400 MHz, and, for simplicity, do not include radio-frequency-dependent noise in the simulations. 
In real PTA data, frequency dependent timing effects such as dispersion measure (DM) variations due to the interstellar medium and solar wind are modeled using wideband or multi-frequency observations and corresponding chromatic noise terms in the timing model. When modeled consistently, we expect to achieve similar results.
For simplicity, we also assume that the data is obtained using only one backend-receiver combination. 
Our simulations contain a single SMBHB merger signal, combined with a gravitational wave background, temporally-uncorrelated noise referred to as ``white'', and temporally-correlated noise referred to as ``red''. 
In this section, we provide the details of our analysis and noise models.

\subsubsection{Pulsar noise}

It is convenient to express pulsar noise in terms of the power spectral density of residuals, $P(f)$, where $f$ is the Fourier frequency of the timing residuals. 
It also corresponds to the frequency of gravitational wave signals. 
White noise (WN) corresponds to the constant power spectral density across frequency, and it is assumed to be known in our simulations. 
Red noise (RN) is more prominent towards the lowest accessible frequency given by the inverse of the observation span, $T_\text{obs}$. 
We assume the power-law power spectral density of red noise \citep{vanHaasterenLevin2009, ColesHobbs2011, ShannonCordes2010}
\begin{equation}
 \label{eq:PSD_rn}
   P(f_j) = \frac{A^2}{12 \pi^2} \left( \frac{f_j}{f_\mathrm{yr}} \right)^{-\gamma} \mathrm{yr}^3, 
\end{equation}
where the amplitude $A$ is in the units of gravitational wave strain at the reference frequency $f_\mathrm{yr} = 1,\mathrm{yr}^{-1}$ and $\gamma$ is the spectral index. 
We simulate and model red noise as a Gaussian process using a Fourier basis. 
In particular, the time series is a sum over $N_f$ sine and cosine terms:
\begin{equation}
 r_\text{red}(t) = \sum_{j=1}^{N_f} \left[ a_j \sin(2\pi f_j t) + b_j \cos(2\pi f_j t) \right].
\end{equation}
Fourier amplitudes $(a_j, b_j)$ are Gaussian random variables with zero mean and the standard deviation determined by $\sqrt{P(f)/ T_\text{obs}}$, an approximation that has been shown to accurately reproduce the target Gaussian process \citep{LentatiAlexander2013, vanHaasterenVallisneri2015, AllenRomano2025, CrisostomivanHaasteren2025}.
In simulated PTA data with red noise, the hyperparameters $(A_{\mathrm{RN}},\gamma_{\mathrm{RN}})$ for intrinsic per pulsar red noise are randomly drawn from uniform distributions over the ranges $[-15,-14]$ and $[1,4]$, respectively.

\subsubsection{Gravitational Wave Background}

In some simulations, we test the robustness of our parameter estimation for SMBHB mergers to the presence of an additional gravitational wave signal.
Namely, the stochastic gravitational wave background (GWB).
It is modeled as a Gaussian process with the same power spectrum in all pulsars and additional inter-pulsar correlations which follow the Hellings–Downs function of an isotropic and unpolarized background \cite{HellingsDowns1983} 
\begin{equation}
\label{eqn:hd}
    \Gamma_{ab} = \frac{1}{2}\delta_{ab} + \frac{1}{2} - \frac{x_{ab}}{4} + \frac{3}{2}x_{ab}\ln x_{ab},
\end{equation}
where $x_{ab} = (1 - \cos\zeta_{ab})/2$, $\delta_{ab}$ is the Kronecker delta function, and $\zeta_{ab}$ is the sky separation angle for a given pair of pulsars.
The power spectral density is
\begin{equation}
 \label{eq:PSD_gwb}
   P_\mathrm{GWB}(f) = \frac{A_{\mathrm{GWB}}^2}{12 \pi^2} \left( \frac{f}{f_\mathrm{yr}} \right)^{-\gamma_\mathrm{GWB}} \mathrm{yr}^3.
\end{equation}
Some of our simulations contain a GWB with $A_\mathrm{GWB} = 10^{-15}$ and $\gamma = 13/3$. 
This choice of a spectral index corresponds to the GWB from a population of adiabatically inspiralling SMBHBs in circular orbits. 
For the purpose of detecting SMBHB mergers, GWB is a source of pulsar-correlated noise.

\subsubsection{Likelihood}

In this work, we perform an analysis of simulated PTA data using Bayesian inference. 
We construct a multivariate Gaussian likelihood. 
The likelihood of the vector of timing residuals $\bm{\delta t}$  is a multivariate Gaussian function \citep{vanHaasterenLevin2009,ArzoumanianBrazier2016,AntoniadisArzoumanian2022}

\begin{equation}
\mathcal{L}(\bm{\delta t} | \bm{\theta}) =
\frac{1}{\sqrt{\det(2\pi \bm{C})}} 
\exp\Big[ 
    -\frac{1}{2} (\bm{\delta t} - \bm{\mu})^\text{T} \bm{C}^{-1} (\bm{\delta t} - \bm{\mu}) 
\Big],
\end{equation}

where $\bm{\mu}$ represents the time series model for an SMBHB merger with null memory. 
Our model parameters are $\bm{\theta}$. 

The contribution of stochastic processes that include noise and the GWB is encoded in the covariance matrix 
\begin{equation}
\bm{C}=\bm{N} + \bm{T} \bm{B} \bm{T}^\text{T}.
\end{equation}
The diagonal white noise matrix $\bm{N}$ has elements $\sigma$. 
We simulate the values $\sigma^2= \sigma_\text{TOA}^2$, and we assume them as fixed parameters. 
Component $\bm{T}$ is the design matrix and $\bm{B}$ is the prior matrix. 
$\bm{T} = [\bm{M}, \bm{F}]$, where $\bm{M}$ is a matrix of partial derivatives of the TOAs with respect to each timing model parameter and $\bm{F}$  is the Fourier design matrix consisting of fourier basis functions that represent long term, time-correlated processes such as red noise and the gravitational wave background. Each component in $\bm{T}$ has a corresponding vector of coefficients, and together these coefficients form a combined vector $b = [\epsilon,a]^\text{T}$ where $\epsilon$ contains timing model corrections and $a$ contains the sine and cosine amplitudes of the Fourier representation of correlated signals. These coefficients are treated as nuisance parameters and are marginalized over in the likelihood. To perform this marginalization efficiently, each group of coefficients is assigned a Gaussian prior with a block-diagonal covariance matrix  $\text{diag}(\bm{B}) = [\bm{\xi},  \bm{\varphi}]$, where $\bm{\xi}$ corresponds to the prior on the timing model parameters which is a broad diagonal prior (values of $10^{40}$) and $\bm{\varphi}$ accounts for temporally correlated processes such as intrinsic spin noise and a stochastic gravitational wave background. The covariance matrix of $\bm{\Phi}$ has elements :
\begin{equation}
\label{eqn:covariance}
    \bm{\Phi}_{(ai),(bj)} = P_{ai}\delta_{ab}\delta_{ij} + \Gamma_{ab} P_{i}\delta_{ij},
\end{equation}
where $a$ and $b$ are pulsar indices, and $i$ and $j$ are frequency indices. $P_{ai}$ is the PSD of the red noise of pulsar $a$ at a frequency bin $i$ (Equation~\eqref{eq:PSD_rn}), $P_i$ is the PSD of the gravitational wave background at the frequency $i$ (Equation~\eqref{eq:PSD_gwb}), and $\Gamma_{ab}$ is the overlap reduction function that determines the degree of this correlation between pulsars $a$ and $b$.

The posterior distribution over parameters is obtained using Bayes' theorem, which reads
\begin{equation}
\mathcal{P}(\bm{\theta} | \bm{\delta t}) = \frac{\mathcal{L}(\bm{\delta t} | \bm{\theta}) \pi(\bm{\theta})}{\mathcal{Z}},
\end{equation}
where $\pi(\bm{\theta})$ is the prior and $\mathcal{Z}$ is the marginal likelihood (evidence) such that
\begin{equation}
\mathcal{Z} = \int \mathcal{L}(\bm{\delta t} | \bm{\theta}) \pi(\bm{\theta}) d\bm{\theta}.
\end{equation}
Model comparison is performed using the Bayes factor, defined as the ratio of evidence of models. 
In cases where models are assigned equal prior odds, this is equivalent to the posterior odds ratio. In this work we do not evaluate the evidences $\mathcal{Z}$ via direct numerical integration. Instead, Bayes factors are computed using the product space (hypermodel) method \citep{HeeHandley2016} implemented in \texttt{enterprise\_extensions}~\citep{enterprise}, in which a model index is sampled jointly with the model parameters and posterior model probabilities are estimated from the relative fraction of samples assigned to each model.

\section{\label{sec:results}Results}

\subsection{Simulated datasets with SMBHB Merger}

We simulate a pulsar timing array consisting of 25 pulsars with 100-ns precision distributed uniformly across the sky. 
The observation spans over 13 years, from MJD 53000 to 58600, with 390 TOAs and observation gaps ranging from a few days to two weeks. 
A large, well-distributed pulsar array improves sky localization of the source and enhances the response to the GW signal.
A noise-free set of time-of-arrival (TOA) measurements is first generated and used as the baseline for constructing multiple datasets for testing signal recovery. 

We simulate two different datasets containing signal from individual SMBHB mergers with observer-frame chirp masses $\mathcal{M}_c = 10^8$ and $10^{10}\,M_\odot$ at luminosity distances of 3 and 100~Mpc, respectively. 
Both binaries are assumed to be equal in mass, circular and non-spinning. They are located at a polar and azimuthal angle of  $(\theta,\phi)=(\pi/3,2\pi/3)$ radians, with a wave polarization of $0$~deg and an inclination of $90^\circ$, and a coalescence time MJD~57000.
The SMBHB merger model includes a continuous inspiral, which introduces quasi-sinusoidal features that can lead to degeneracies across parameters such as sky location, chirp mass, luminosity distance, and merger time. 
These combinations of degenerate parameters, in turn, can produce multiple local maxima in the likelihood. 
This situation is analogous to challenges encountered in continuous gravitational wave searches, where multimodal posteriors have been noted and addressed using specialized sampling algorithms~\citep{PetiteauBabak2013}. 
To address this multimodal likelihood surface, we use a parallel tempering Markov chain Monte Carlo (PTMCMC) sampler~\citep{EllisvanHaasteren2019}. 

In PTMCMC, multiple chains are run simultaneously at different temperatures, $T$ characterized by the inverse temperature parameter $\beta = 1/T$. The posterior distribution at a given temperature is
\begin{equation}
    \label{eq:temperature}
    P_\beta(\theta \mid \delta t ) \propto \mathcal{L}(\delta t  \mid \theta)^\beta \, \pi(\theta)
\end{equation}
where $\mathcal{L}(\delta t  \mid \theta)$ is the likelihood and $\pi(\theta)$ is the prior. As $\beta$ goes from 0 to 1, we effectively move from the prior-only distribution ($\beta=0$) to the posterior ($\beta=1$). $\beta< 1$ flattens the posterior, reducing the relative heights of peaks and allowing chains to explore the parameter space more freely. This exchange between temperatures improves mixing and convergence. 
We use 20 geometrically spaced temperatures, which we find sufficient to recover posteriors robustly for signals with both oscillatory and memory components.

To test the robustness of parameter estimation to different noise conditions, we construct two types of simulated datasets: 
\begin{enumerate}
    \item Simulation \texttt{wn+rn+mem}: white noise, intrinsic per pulsar red noise whose hyperparameters $(A_{\mathrm{RN}},\gamma_{\mathrm{RN}})$ are drawn from the uniform distributions $[-15,-14]$ and $[1,4]$, respectively, and SMBHB merger.
    \item Simulation \texttt{wn+rn+gwb+mem}: white noise, intrinsic per pulsar red noise, and a stochastic gravitational wave background with Hellings-Downs correlations. 
\end{enumerate}

Each SMBHB merger is simulated on both noise realizations, producing a total of four datasets spanning both the masses and corresponding distances. After simulating the datasets, we perform parameter estimation. 
The parameters and priors used for this analysis are shown in Table \ref{tab:priors}. 
We fix the white noise parameters to their simulated values to reduce dimensionality and focus the sampling on the recovery of the signal and red noise components. 
We adopt log-uniform priors on the chirp mass and luminosity distance. These choices correspond to scale-invariant, weakly informative priors appropriate for parameters spanning multiple orders of magnitude. Our goal is to demonstrate the recovery performance of the waveform model without imposing strong astrophysical assumptions on the underlying population.

\begin{table*}
\caption{\label{tab:priors}Details of the prior distributions of various parameters used in the Memory Signal Recovery analysis.}
\renewcommand{\arraystretch}{1.3}
\begin{ruledtabular}
\begin{tabular}{ccc}

 \textbf{Parameter}&\textbf{Prior}&\textbf{Description}\\ \hline
  \multicolumn{3}{c}{\textbf{Intrinsic Pulsar Red Noise (RN)}}\\ 
 $\log_{10}A_\text{RN}$   & $\mathcal{U}(-17,\,-12)$ & Log–10 amplitude of Red noise \\ 
 $\gamma_\text{{RN}}$   & $\mathcal{U}(1,\,7)$ & Spectral index of Red noise\\ \hline
 \multicolumn{3}{c}{\textbf{Gravitational Wave Background (GWB)}}\\ 
 $\log_{10}A_\text{{GWB}}$   & $\mathcal{U}(-20,\,-6)$ & Log–10 amplitude of GWB\\
 $\gamma_\text{{GWB}}$   & $\mathcal{U}(1,\,7)$ & Spectral index of GWB\\ \hline
 \multicolumn{3}{c}{\textbf{SMBHB Merger with Null Memory}}  \\ 
 $\log_{10}\mathcal{M}$~[$M_\odot$]   & $\mathcal{U}(8,\,12)$ & Observer-frame Chirp Mass  \\ 
    $\log_{10}D_{L}$~[Mpc]   & $\mathcal{U}(0,\,5)$ & Luminosity Distance  \\ 
 $q$   & $\mathcal{U}(1,\,7)$ & Mass ratio \\
 $\psi$ & $\mathcal{U}(0,\,\pi)$ & Polarization  \\
 $\cos\theta$           & $\mathcal{U}(0,\,\pi)$   & Cosine of Polar angle       \\
 $\phi$             & $\mathcal{U}(0,\,2\pi)$  & Azimuthal angle   \\
  $\iota$             & $\mathcal{U}(0,\,\pi)$  & Inclination angle   \\
 $t_\text{merge}$~[MJD]            & $\mathcal{U}(53000,\,58600)$  & Time of merger\\ 
\end{tabular}
\end{ruledtabular}
\end{table*}

\begin{figure*}
  \centering
  \begin{subfigure}{0.45\textwidth}
    \includegraphics[width=\textwidth]{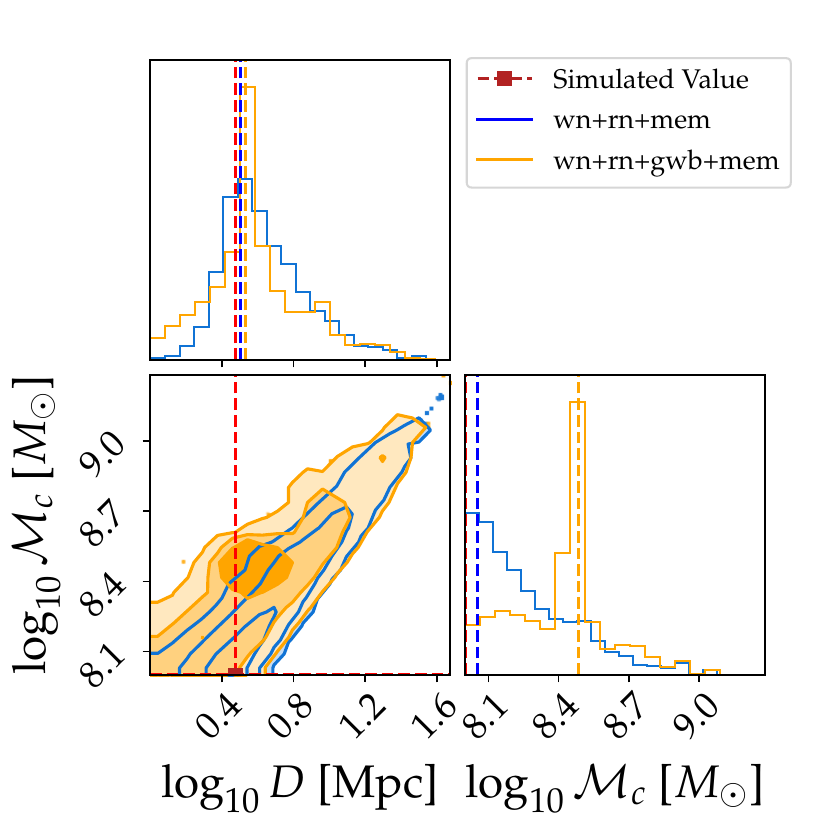}
    \caption{}
    \label{fig:M8}
  \end{subfigure}
  \hfill
  \begin{subfigure}{0.45\textwidth}
    \includegraphics[width=\textwidth]{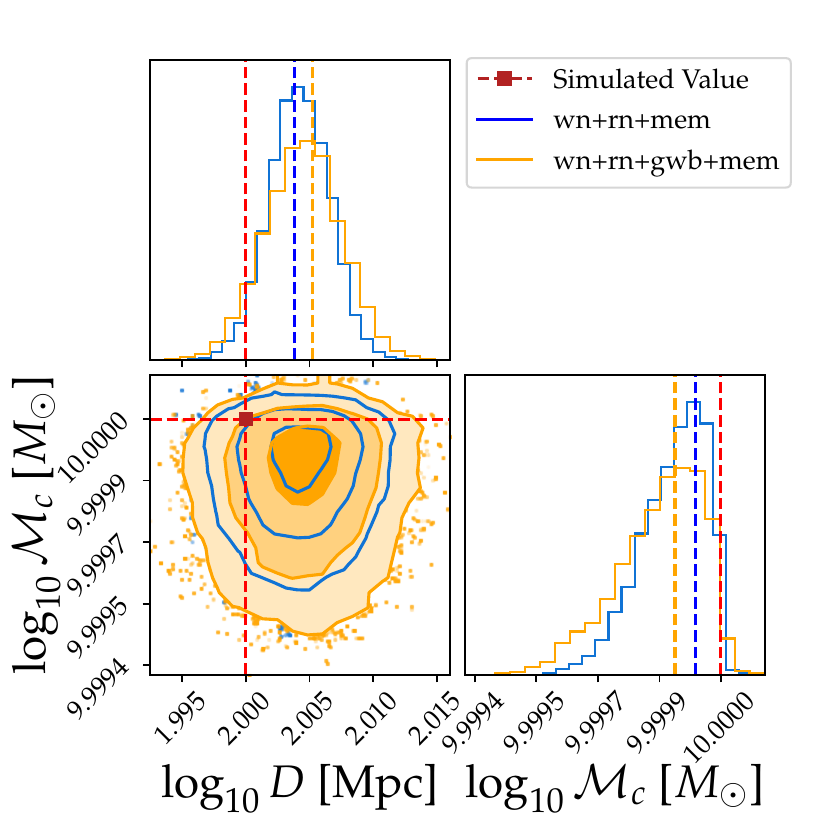}
    \caption{}
    \label{fig:M10}
  \end{subfigure}
  \caption{\justifying The figure illustrates the posterior distributions from our simulation recovery studies with 25 pulsars. Simulation studies were conducted with 2 types of datasets - \texttt{wn+rn+mem} and \texttt{wn+rn+gwb+mem} and the contours are shown in blue and orange respectively. Three subfigures show a different SMBHB merger.
  (a) A merger with $\mathcal{M}_c = 10^{8} M_{\odot}$ and $D_\text{L} = 3$ Mpc and,
  (b) A merger with $\mathcal{M}_c = 10^{10} M_{\odot}$ and $D_\text{L} = 100$ Mpc. Maximum-\textit{a-posteriori} (MAP) values are indicated by dotted lines in the respective contour colors (blue and orange), and the true simulated values are shown as red dotted lines.
  }
  \label{fig:mass_dist_corner}
\end{figure*}

Figure \ref{fig:mass_dist_corner} shows the joint posterior distributions of chirp mass and luminosity distance for the four different datasets. 
For the lower-mass system shown in Fig.~\ref{fig:M8}, analyzed without a simulated stochastic gravitational-wave background, both the chirp mass and luminosity distance are recovered with the injected values lying within the 68\% marginal credible intervals. When a stochastic background is included, the marginal posterior for the chirp mass shifts toward higher values and the maximum-\textit{a-posteriori} (MAP) estimate is displaced from the simulated mass. However, the injected parameters remain consistent with the data, lying within the joint $2–3\sigma$ credible region of the 
posterior. This behavior is a consequence of the strong mass–distance degeneracy characteristic of memory-dominated PTA signals since the GW memory strain $h_{\text{mem}} = \mathcal{M}_c/D_{\text{L}}$. The simulated chirp mass in Fig.~\ref{fig:M8} lies close to the lower prior bound, which was chosen based on PTA sensitivity considerations. Systems with smaller masses would produce memory signals that are not expected to be identifiable or constrained with the current PTA sensitivity. For this simulation, the memory strain amplitude is comparatively small $h_{\text{mem}} \approx 4 \times 10^{-14}$ and dominated by low-frequency power. As a result, the inclusion of a correlated low-frequency process such as a stochastic gravitational-wave background increases the effective covariance at long timescales, broadening the posterior and allowing the MAP estimate to shift along the mass–distance degeneracy without excluding the true parameters.
In contrast, for the higher-mass system shown in Fig.~\ref{fig:M10} ($10^{10} M_\odot$), the projected memory amplitude is significantly larger ( $h_{\text{mem}} \approx 1 \times 10^{-13}$), leading to tighter constraints. In this case, both the chirp mass and luminosity distance are recovered within the 68\% marginal credible intervals in both noise configurations, and the MAP estimates remain close to the injected values.
The posterior distribution for all parameters of the SMBHB merger are given in Appendix.

\begin{figure*}
  \centering
  \begin{subfigure}{0.45\textwidth}
    \includegraphics[width=\textwidth]{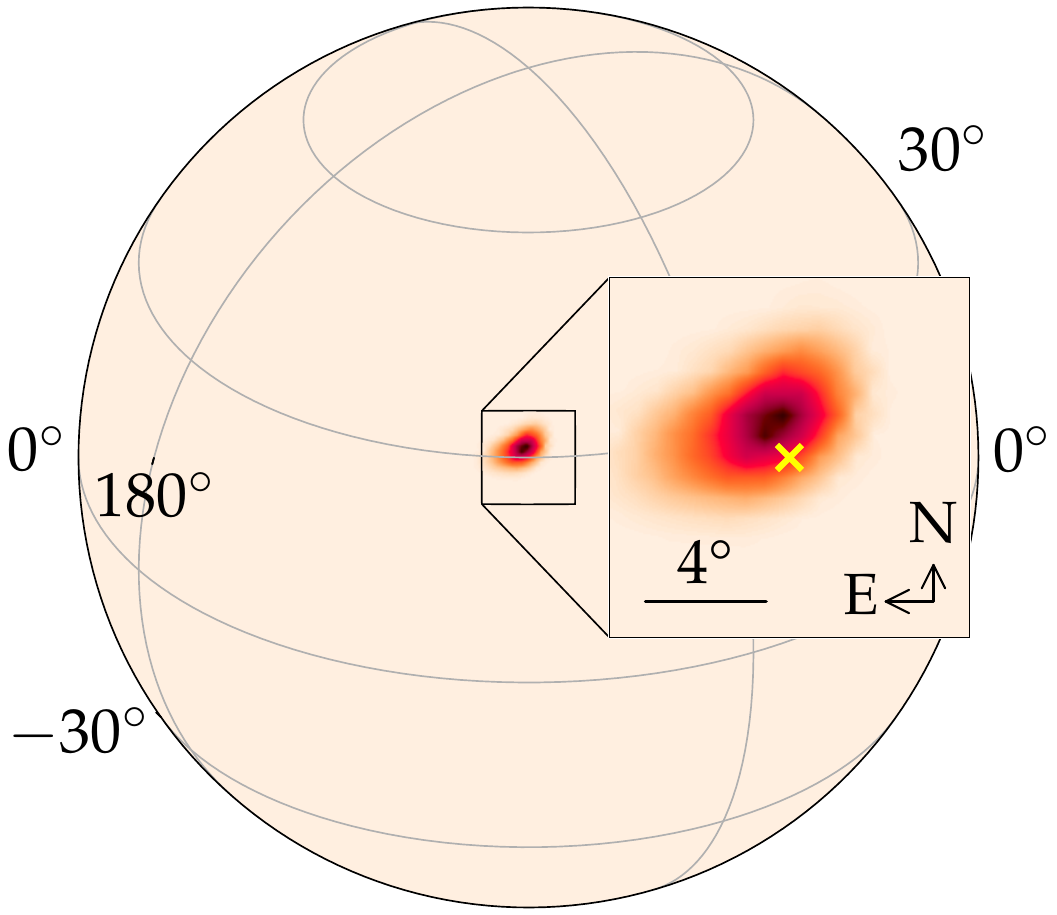}
    \caption{}
    \label{fig:M8_skymap}
  \end{subfigure}
  \hfill
  \begin{subfigure}{0.45\textwidth}
    \includegraphics[width=\textwidth]{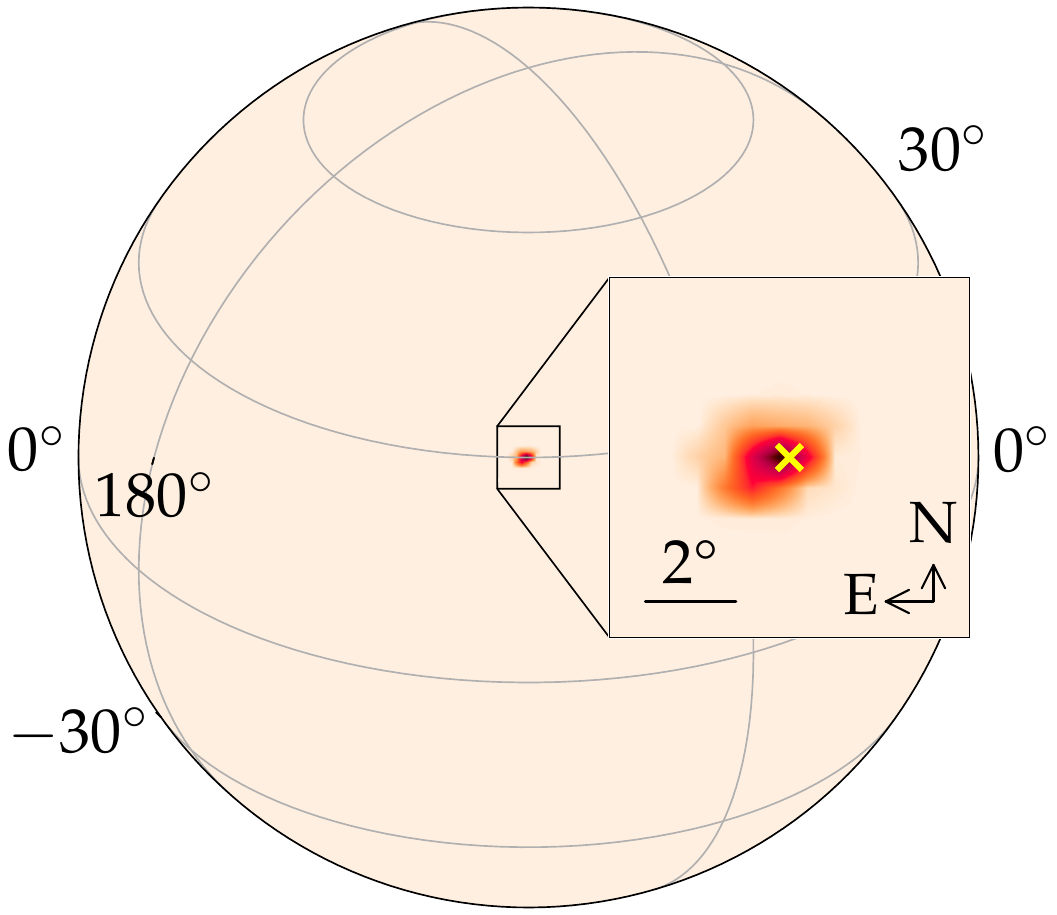}
    \caption{}
    \label{fig:M10_skymap}
  \end{subfigure}
  \caption{\justifying Sky maps of the posterior distributions of simulated supermassive black hole binary mergers, shown in equatorial coordinates (RA, Dec in degrees). The color scale represents the relative posterior probability density derived from our HEALPix projections of the samples, while the yellow cross marks the true (simulated) sky location of the source. Each square projection shows a zoomed region around the source, with the scale bar indicating the angular width on the sky. The three subfigures correspond to mergers with increasing chirp mass and luminosity distance: (a) $\mathcal{M}_c = 10^8 M_\odot$, $D_L = 3$ Mpc. ; (b) $\mathcal{M}_c = 10^{10} M_\odot$, $D_L = 100$ Mpc. As the chirp mass increases, the posterior uncertainty in sky localization decreases, demonstrating improved measurement precision for more massive mergers.}
  \label{fig:skymaps}
\end{figure*}

To illustrate the sky localization performance of our simulated supermassive black hole binary (SMBHB) mergers, we projected the posterior samples onto HEALPix sky maps in equatorial coordinates (RA, Dec in degrees). This is shown in Figure~\ref{fig:skymaps}. The color scale in each map represents the relative posterior probability density, and the yellow cross marks the true (simulated) source location. Each map includes a zoomed inset that highlights the region around the source, with a scale bar indicating the angular width of the zoom. As the chirp mass increases, the precision of sky localization improves noticeably. This trend reflects the stronger gravitational wave signal associated with more massive mergers, which allows the parameter estimation algorithms to constrain the source position more accurately. The insets also visually emphasize how the uncertainty region shrinks with mass, showing that even at larger distances, more massive binaries can be localized more precisely than lighter, closer systems. 

Additionally, we perform a model selection using product space method between the memory and the noise-only model to evaluate the detection confidence. In the case of both \texttt{wn+rn+mem} and \texttt{wn+rn+gwb+mem} datasets, all posterior samples favor the memory model, allowing us to place a conservative lower bound on the Bayes factor of $\ln \mathcal{B} > 10$. Throughout this work, model preference and detectability are quantified using Bayes factors rather than matched-filter signal-to-noise ratio.

\subsection{Limitations of the Memory Burst Approximation and Its Impact on Observable Signals}

In Figure 1 of ~\citet{Favata2009}, a comparison is presented between the nonlinear memory waveform computed using an Effective-One-Body (EOB) and Minimum Waveform model (MWM). Favata reports that the EOB-based model underestimates the final memory strain offset by $\approx27\%$ relative to the NR calculations, even after calibration. This discrepancy arises due to simplifications in the analytical model, particularly during the merger and ringdown phases, as well as the omission of higher-order radiative modes.

The analytic expression for memory strain used by \citet{PshirkovBaskaran2010} is based on a further simplified approximation derived from the nonlinear memory formula presented in \citet{Favata2009}. In particular, they approximate the radiated energy during the merger as a fixed mean value and assumes an equal-mass, non-spinning binary on a circular orbit. 

As a result, the predicted strain is systematically larger than what would be obtained from physically realistic SMBHB models. This leads to an overestimation of the memory amplitude or equivalently, an underestimation of the luminosity distance or overestimation of the source mass when interpreting the observed data using this model.

Due to these limitations, the memory burst model does not provide a reliable mapping between the observed signal and the underlying parameters of the SMBHB. Even if the final strain offset is calibrated correctly, the second limitation of the burst model, which is the absence of gradual memory growth, persists. This has important observational consequences, particularly in the context of pulsar timing array (PTA) searches.

PTA analyses are not sensitive to the gravitational wave memory time series in full.
Instead, they are sensitive to the \textit{post-fit residuals}.
That is, the signal remaining after subtraction of the best-fit pulsar timing model. 
This subtraction typically includes a polynomial fit for the spin frequency and its time derivatives. Any signal that mimics these trends can be partially absorbed, reducing the amplitude of the observable signature.

\begin{figure*}
  \centering
  \begin{subfigure}{0.32\textwidth}
    \includegraphics[width=\textwidth]{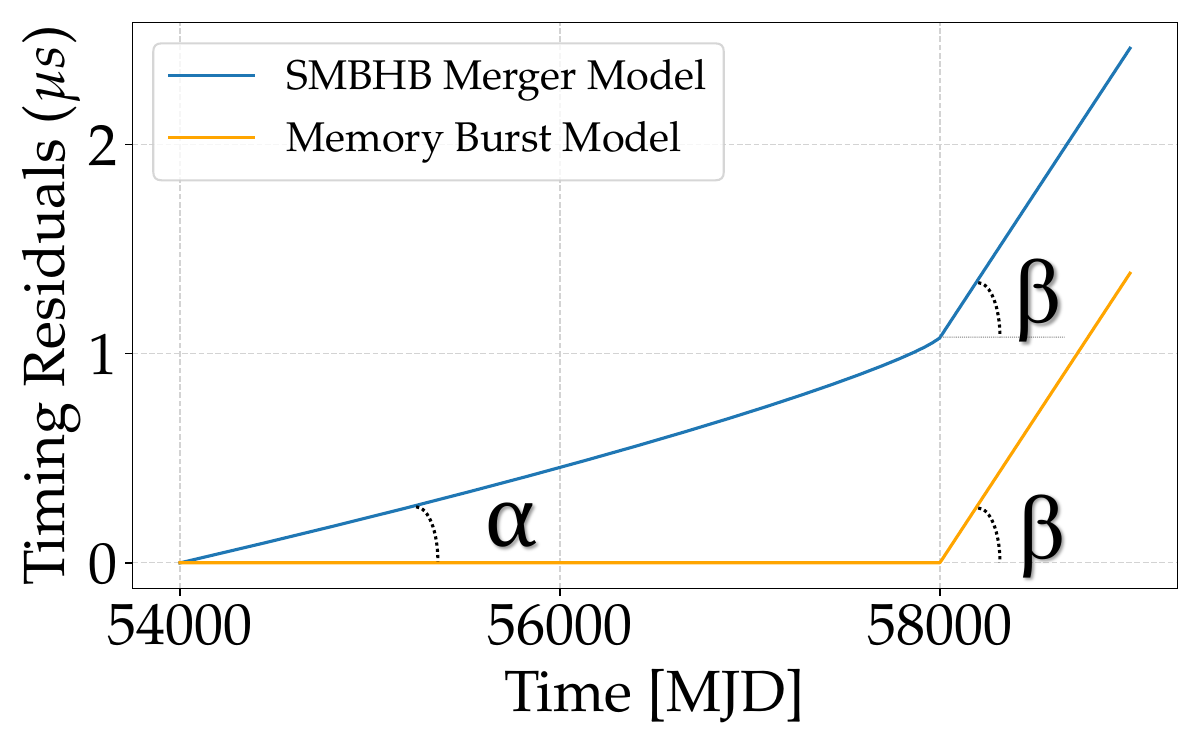}
    \caption{}
    \label{fig:comp}
  \end{subfigure}
  \hfill
  \begin{subfigure}{0.32\textwidth}
    \includegraphics[width=\textwidth]{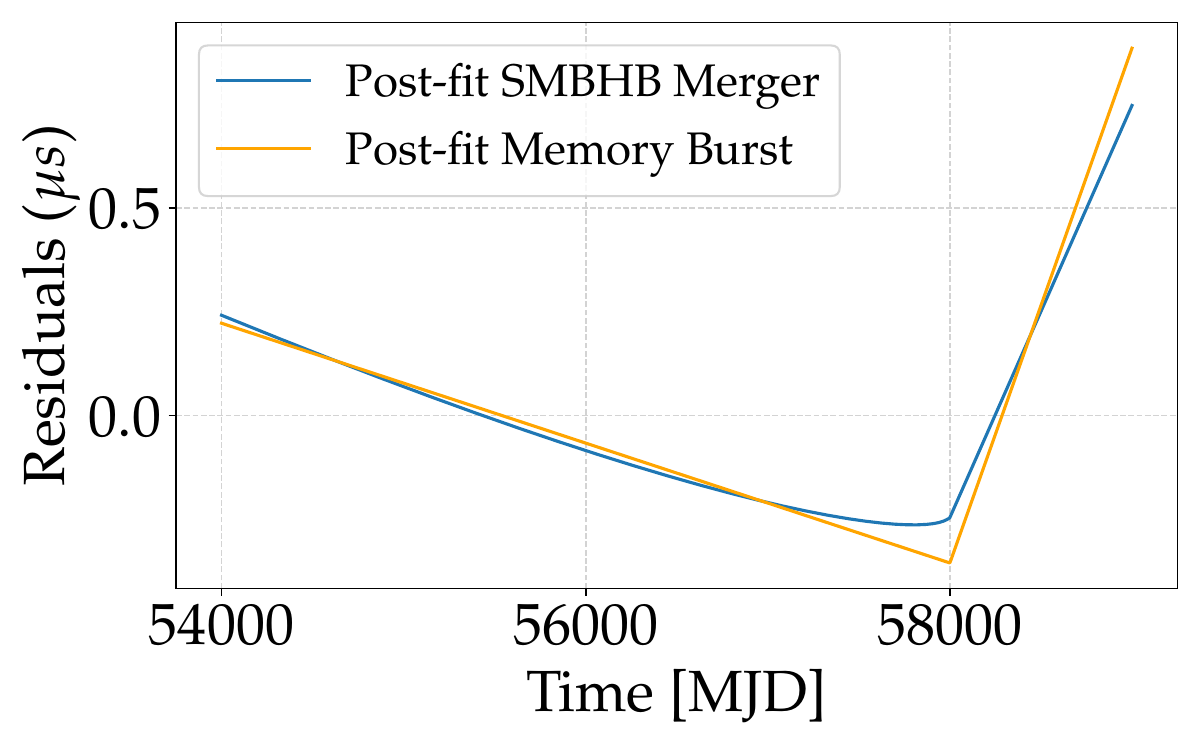}
    \caption{}
    \label{fig:postfit}
  \end{subfigure}
  \hfill
  \begin{subfigure}{0.32\textwidth}
    \includegraphics[width=\textwidth]{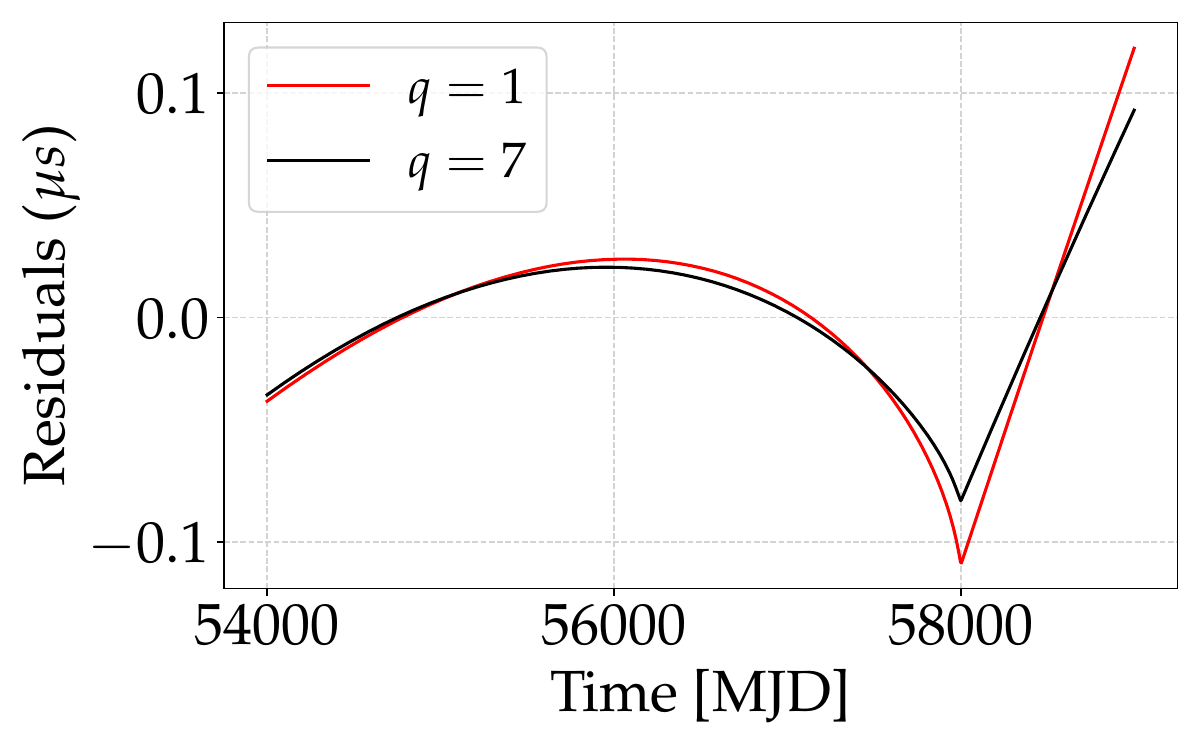}
    \caption{}
    \label{fig:diff}
  \end{subfigure}
  \caption{\justifying Timing residuals from a pulsar located at (ra,dec) =  ($258.4564^{\circ}$, $7.7937^{\circ}$) induced by a gravitational wave from a SMBHB merger with $\mathcal{M}_c = 10^{10} M_\odot$, $D_L = 100$ Mpc and $q=1$ located at ($0^\circ$ ,$0^\circ$). (a) Prefit timing residuals for memory burst model (orange) and the SMBHB merger model (blue). Both models are adjusted to have the same final memory offset. The SMBHB model shows a gradual pre-merger rise at angle $\alpha$, whereas the burst model shows an abrupt change at angle $\beta$.
  (b) Post-fit residuals after subtraction of a linear spindown model. The observable residual bump for the SMBHB model (blue) is suppressed compared to the burst model (orange), due to pre-merger curvature being partially absorbed by the fit. For the equal-offset case, the post-fit RMS difference between the models is $0.399\,\mu\mathrm{s}$.
  (c) Difference in post-fit timing residuals between the burst model and SMBHB model for varying mass ratios. The discrepancy is largest for the equal-mass case ($q=1$) and decreases for more unequal-mass binaries. The RMS difference drops by $\sim 0.1\,\mu\mathrm{s}$ when the mass ratio increases from $q=1$ to $q=7$.
  }
  \label{fig:model_comparison}
\end{figure*}

The gravitational wave memory signal from a realistic SMBHB merger builds up gradually over the inspiral and coalescence, in contrast to the instantaneous step-function approximation assumed in the standard burst model. Even when both burst model and SMBHB merger model are calibrated to reach the same final strain offset, their evolution in time (particularly around the merger) leads to important differences in the observable residuals after subtraction of the pulsar timing model. For the comparison in Figure~\ref{fig:model_comparison}, we adjust the final strain offset to be the same across both models for a SMBHB binary of $\mathcal{M}_c = 10^{10} M_\odot$ and $D_L =100$ Mpc. This isolates the effect of the signal's shape rather than its amplitude. 

Figure~\ref{fig:comp} compares the pre-fit residuals produced by the memory burst model (orange) and the full SMBHB merger model (blue) for the same simulated binary. Throughout this section, we treat the SMBHB waveform which includes the pre-merger memory buildup as the most accurate available representation of the true gravitational-wave memory signal. Both models exhibit a monotonic post-merger rise corresponding to the final memory offset, while the SMBHB waveform additionally shows a gradual pre-merger increase associated with the inspiral-driven buildup of memory. In contrast, the burst model approximates this growth as an instantaneous ramp at the merger epoch.

After subtracting the pulsar spin down and its derivatives, the post-fit residual differs significantly between the two waveform models for the same SMBHB. This is depicted in Figure~\ref{fig:postfit}. Since the SMBHB waveform contains a slow pre-merger rise extending over a long timescale, a significant fraction of its signal is absorbed by the timing-model fit, whereas the burst approximation concentrates the entire memory contribution near the merger time and is therefore less affected. As a result, even when both models are normalized to produce the same final memory offset, the observable post-fit residuals differ.

To quantify this mismatch in the PTA observable, we compute the noise-agnostic root-mean-square (RMS) of the difference between the post-fit residuals. For our fiducial source ($\mathcal{M}_c=10^{10}M_\odot$, $D_L=100$\,Mpc, $q=1$), enforcing the same final memory offset yields a post-fit RMS difference of $0.399\, \mu \mathrm{s}$. Allowing the burst-model strain amplitude to vary in order to minimize this post-fit RMS reduces the mismatch to $0.26\, \mu \mathrm{s}$. In this best-fit configuration, the computed burst model strain amplitude is larger by 11.8\% compared to the amplitude from the SMBHB merger model. Interpreting this best-fit burst amplitude as a physical memory step leads to biased source parameters. Fixing the simulated luminosity distance and inferring the chirp mass from the adjusted burst amplitude (Equation~\eqref{eq:strain_mass_dist} results in 37.5\% discrepancy in the observer-frame chirp mass relative to the injected value; conversely, fixing the chirp mass and inferring the distance produces a 62.2\% bias. Thus, even when optimally tuned to match the post-fit residuals of the true SMBHB signal, the simplified burst approximation yields biased parameter estimates and leaves a residual mismatch that ultimately reduces sensitivity. 

Figure ~\ref{fig:diff} illustrates the difference between the post-fit residuals of the two models. These differences are primarily driven by the gradual pre-merger memory buildup present in the SMBHB waveform but absent in the burst approximation, and therefore depend on the binary’s mass, mass ratio, distance, and spin (we assume non-spinning binaries here). As the mass ratio becomes more unequal, the memory builds more steeply during the inspiral phase. We find that the mismatch is largest for equal-mass systems and decreases for more unequal mass ratios. For example, the RMS difference between the residuals decreases by approximately $0.1 \,\mu s$ as the mass ratio increases from $q = 1$ to $q = 7$.

This bias is not a mere modeling artifact. 
It has real implications for data analysis. If a PTA search uses an idealized burst template while the true signal evolves gradually, as in realistic SMBHB mergers, the recovered signal may appear artificially strong. This leads to incorrectly inferred source parameters, such as an underestimated luminosity distance or overestimated strain. This highlights the importance of accurate waveform modeling in PTA-based memory searches.

\subsection{Null Hypothesis Tests and Upper Limits}

We simulate a dataset containing only pulsar-specific white and red noise to demonstrate our ability to correctly place upper limits and to ensure that we do not detect any spurious evidence for SMBHB mergers. 

We use the same PTA configuration as in the strong signal simulation study. Namely, a 25-pulsar array with uniform sky coverage, 13-year observing span but with 1000 TOAs and $\sim$ 6-day observation cadence.
We then perform a model selection between the correct null hypothesis and the two signal hypotheses.
Signal hypotheses additionally include the memory burst and the SMBHB merger signal, respectively.
We find log Bayes factors of $-0.375$ for the memory burst model and $0.1$ for the SMBHB merger model. These values indicate no significant evidence for a memory signal in either case, as expected.

\begin{figure}[htbp]
  \centering
  \includegraphics[width=\columnwidth]{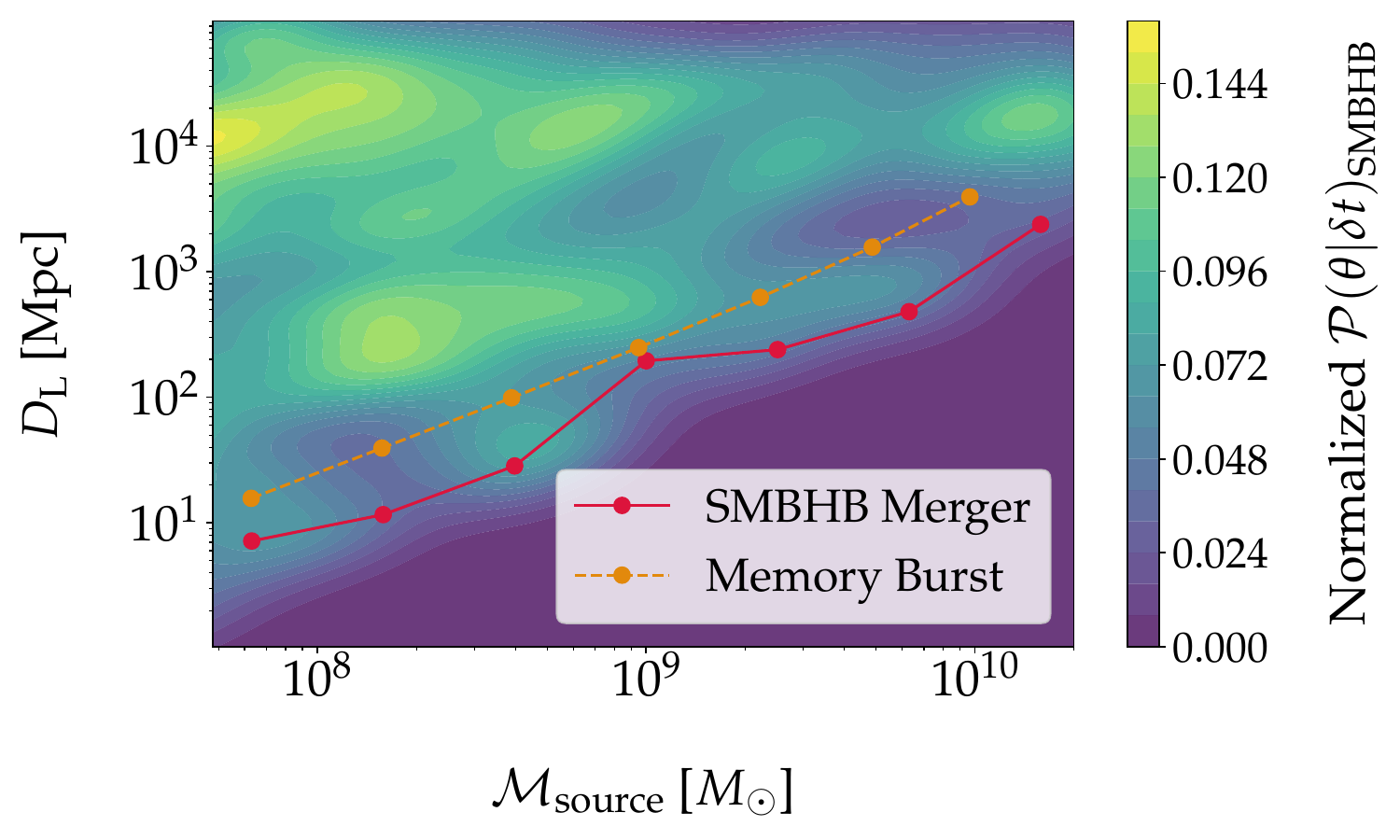}  
  \caption{\justifying
  {Lower limits on luminosity distance as a function of chirp mass in the signal-free simulated data. The filled contours show the marginalized posterior density in the $\mathcal{M}$–$D_L$ plane using the full SMBHB merger waveform model. The red curve represents the exclusion boundary below which a source would have been detectable with $95\%$ confidence. The orange curve shows the corresponding lower limit derived from the $95\%$ upper limit on memory strain amplitude using the memory burst model, converted to distance via Equation~\eqref{eq:strain_mass_dist}. 
  }}
  \label{fig:dist_ll}
\end{figure}

Figure~\ref{fig:dist_ll} summarizes the limits on the source frame chirp mass and the luminosity distance based on the SMBHB merger model and the memory burst model applied to the simulated data. 
For the SMBHB merger model, we find limits on both the chirp mass and luminosity distance. 
The filled contours represent the marginalized posterior density in the chirp mass–luminosity distance plane for the SMBHB model in the noise-only data. 
The red curve shows $95\%$ lower limits on the luminosity distance for fixed bins in chirp mass, representing minimum distances to which our analysis would have been sensitive for a signal of a given chirp mass. 
For the memory burst model, we obtain a $95\%$ upper limit on the strain amplitude of a gravitational-wave memory event. 
To translate this strain limit into a distance constraint, we adopt the analytic scaling of \citet{PshirkovBaskaran2010}, rewritten in terms of source-frame chirp mass and luminosity distance. For an equal-mass binary ($q=1$) the memory strain can be expressed as

\begin{equation}
\label{eq:strain_mass_dist}
h_\text{mem} \approx 1.13 \times 10^{-15} \left(\frac{\mathcal{M}_\text{source} (1+z)}{10^8 M_\odot}\right) \left(\frac{1~\text{Gpc}}{D_{L}}\right),
\end{equation}

where $D_{L}$ is the luminosity distance and $z$ is the corresponding cosmological redshift assuming a flat $\Lambda_\text{CDM}$ model. While the signal observed at Earth depends on the redshifted (observer-frame) chirp mass $\mathcal{M}_\text{obs} = \mathcal{M}_\text{source}(1+z)$, we report results in the source frame. This provides a straightforward mapping between the measured strain upper limit and a minimum allowed luminosity distance to a hypothetical source, assuming a mass ratio of one.
The orange curve shows the corresponding limit derived using the memory burst model via Equation~\eqref{eq:strain_mass_dist}. 

The difference in the exclusion boundaries between the two models reflects the modeling limitations discussed in the previous section. The memory burst model overestimates the final memory offset due to simplified assumptions such as EOB approximations and no gradual pre-merger buildup. These approximations cause the strain amplitude to appear larger than it would be in a realistic SMBHB merger scenario. As a result, the burst model places artificially stringent constraints on the luminosity distance, leading to higher $D_L$ limits than the SMBHB merger model. This visual difference, shown in Figure~\ref{fig:dist_ll}, illustrates that using the burst approximation can misrepresent the PTA's true sensitivity and potentially bias population-level interpretations based on upper limit studies.

\section{\label{sec:discussion}Discussion}

The method presented in this work represents a major step forward in the search for gravitational wave memory from SMBHBs using pulsar timing arrays. 
It provides several key advantages over earlier approaches that focused solely on memory burst detection. 
In what follows, we highlight the implications of our findings and outline the next steps.

\paragraph{Improvement upon the memory burst model.} 
The SMBHB merger model reduces false positives and improves parameter estimation compared to the simplified memory burst model. Its detailed time structure also helps distinguish real signals from glitches or noise. The memory burst model, by assuming an instantaneous step and simplified energy estimates, tends to overestimate the strain amplitude of realistic binaries. This can lead to biased interpretations such as overestimating the mass of the binary and highlights the importance of using physically consistent waveforms in PTA memory searches to infer accurate astrophysical parameters from the observed signal.

\paragraph{Spinning SMBHBs.}

At present, our analysis is restricted to non-spinning, circular SMBHBs. The package \texttt{NRHybSur3dq8\_CCE} supports aligned-spin binaries. Previous studies have shown that spin can significantly affect the memory amplitude. 
In particular, the final memory offset of strain for maximally aligned spins ($a_1 = a_2 = 1$) is by a factor of $2.5$ larger than for a non-spinning binary~\citet{PollneyReisswig2011}. 
This enhancement arises because higher-spin systems radiate more energy through gravitational waves, which in turn sources a larger null memory. Thus, considering spinning binaries may improve the odds of observing a merging SMBHB.

\paragraph{Eccentric SMBHBs.}

Eccentricity is another important factor not addressed in the present work. 
\citet{Favata2011} presents a derivation of the null memory from binaries on elliptical orbits and shows that the memory flux becomes sharply peaked near each pericenter passage as eccentricity increases. 
This differs from the smoother, continuous buildup of memory in circular binaries. 
While the total memory evolves toward the circular case as the orbit circularizes, systems that maintain eccentricity during their evolution may show a different memory accumulation profile. 
Figure~9 in Ref.~\cite{Sesana2010} shows that a non-negligible fraction of SMBHBs may have significant eccentricities. 
Therefore, it is of interest to search for merging eccentric SMBHBs. 

\paragraph{Prospects for Electromagnetic Counterparts Enabled by the SMBHB Merger Model.}

Electromagnetic (EM) counterparts to supermassive black hole binaries (SMBHBs) have been studied theoretically using both semi-analytic models and general relativistic simulations, predicting a range of possible signatures: precursors in accretion signatures, variable jets, circumbinary disk effects, or flares associated with the final coalescence~\citep{GutierrezCombi2022,BogdanovicMiller2022}. 
These multiwavelength signatures offer rich insights into the binary environment, host galaxy, and physical parameters. 
When applied to future PTA experiments with increased sensitivity, the SMBHB merger model may allow EM follow-up and multi-messenger observations. 
Current PTAs release data once every few years, but EM follow-up of SMBHB mergers requires more frequent data combinations and analysis, updating the TOAs on a timescale of months instead of years, and tracking the evolving significance of the signal could allow.
Once a detectable SMBHB is close to merger, it is possible to increase the frequency of observations and prepare alerts to EM observatories.
This may open up the possibility of detecting counterpart signatures such as disk shocks, jet reconfiguration, or transient EM flares, potentially observable across multiple bands~\citep{GutierrezCombi2022,BogdanovicMiller2022}. 
It may be of interest to investigate this possibility further in future work.

\paragraph{New window onto SMBHBs.} 
Our search for merging SMBHBs with the NR waveform model with null memory opens a new window onto SMBHBs for PTAs. 
The other two windows are the existing searches for stochastic background from SMBHBs and for continuous gravitational waves from SMBHB sources that stand out from the background.

\begin{figure}[htbp]
  \centering
  \includegraphics[width=\columnwidth]{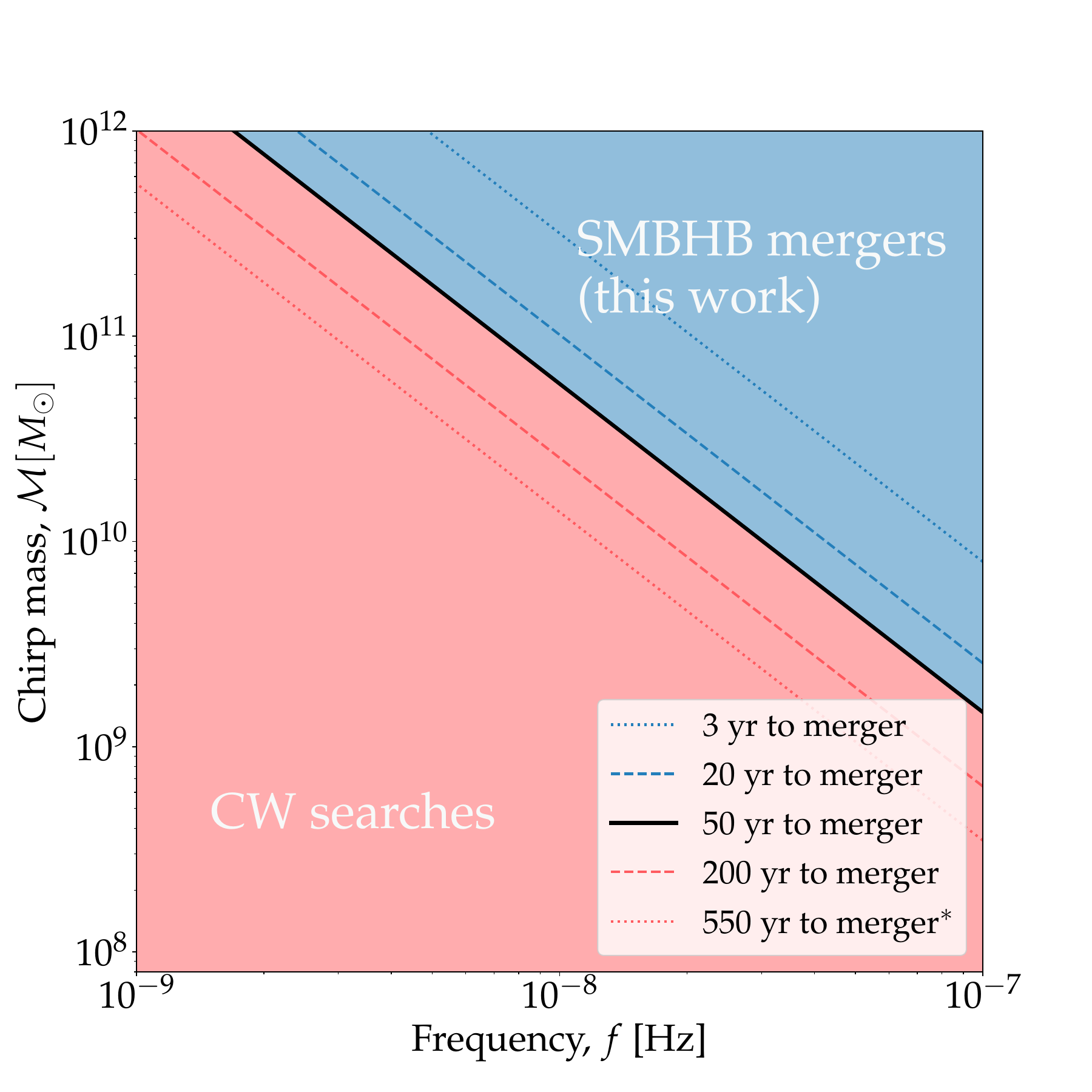}  
  \caption{\justifying
  $\mathcal{M}$ - $f$ parameter space of (1) continuous wave (CW) searches for SMBHBs with PTAs and (2) the SMBHB merger searches that we introduce in this work. Diagonal lines indicate fixed times to the SMBHB merger. 
  The line marked by an asterisk in the legend, corresponding to $550$ years to merger, indicates a case where a signal will slowly drift out of the frequency bin on a timescale of PTA observations in the next decade~\cite{CardinalTremblayGoncharov2025}. 
  }
  \label{fig:parameter_space}
\end{figure}

We compare the parameter-space coverage of CW searches and our merger search in Figure~\ref{fig:parameter_space}.

Diagonal lines of equal time to merger are constructed based on the relation~\citep{AbbottAbbott2017} 
\begin{equation}
    f^{-8/3}(t) = \frac{(8 \pi)^{8/3}}{5} \bigg( \frac{G \mathcal{M}}{c^3} \bigg)^{5/3} (t - t_\text{c}),
\end{equation}
where $G$ is Newton's constant, $c$ is the speed of light, $t_\text{c}$ is the time of coalescence. This relation corresponds to the leading-order quadrupole approximation for quasi-circular binaries; higher post-Newtonian corrections and orbital eccentricity would modify the detailed frequency evolution but do not affect the qualitative parameter-space coverage illustrated in Fig.~\ref{fig:parameter_space}.
Continuous gravitational wave searches typically assume a non-evolving frequency, which is valid up to about $3 \times 10^{-8}$~Hz for a heavy SMBH with $\mathcal{M}=10^{10}~M_\odot$. 
At higher frequencies, the signal will merge on a timescale comparable to that of the PTA observation. 
Moreover, even for an SMBHB that will merge in $550$~years, illustrated by the bottom dotted line, the signal is expected to drift out of the frequency bin. 
This is shown in the targeted search for the SMBHB candidate 3C~66B~\cite{CardinalTremblayGoncharov2025}. 
If the candidate is a real SMBHB, it will also merge in about $550$~years, and the signal may drift out of the frequency bin in the foreseeable future. 
In such regimes, conventional CGW searches lose sensitivity, whereas our method naturally captures both frequency evolution, the late inspiral, and the final merger. 
The gravitational-wave memory contribution is not strongly degenerate with CGWs,even though both are included in the model, as memory produces a monotonic, non-oscillatory signature in timing residuals, while CGWs generate quasi-sinusoidal signals that retain a distinct morphology after timing-model fitting. The presence of additional unmodelled CGW sources in the data would be effectively absorbed into the stochastic background, or be modeled explicitly if individually resolvable. In either case, such signals are expected to broaden posterior distributions rather than mimic or bias the memory parameters. 

\paragraph{Impact of noise systematics and real PTA configurations.}
Chromatic noise processes such as dispersion measure variations, solar wind effects, and other frequency-dependent systematics are not degenerate with the gravitational wave memory signals, which produce an achromatic, non-oscillatory signature in timing residuals. When modeled consistently, such chromatic processes would not bias the recovered memory parameters. Similarly, non-ideal PTA configurations with anisotropic sky distributions and pulsars spanning a wide range of timing precision are accommodated within the PTA likelihood framework and primarily affect sensitivity rather than qualitative inference. The feasibility of applying this method to real PTA data, including sub-banded TOAs, complex noise models, and large data volumes, has been demonstrated in our companion search for gravitational wave memory using EPTA and PPTA datasets with this waveform model \citep{TomsonGoncharov2026}. While future improvements in computational efficiency and noise modeling will be important for next-generation datasets, these challenges do not limit the applicability of the method to current PTA observations.

\section{\label{sec:conclusion}Conclusions}

In this work, we presented a physically complete waveform model for detecting gravitational-wave memory signals from supermassive black hole binaries (SMBHBs) in pulsar timing array (PTA) data, extending beyond the conventional burst-with-memory search strategies. 
Our method models the full memory waveform inspired by numerical relativity, allowing us to directly explore a broader astrophysical parameter space including binary mass, luminosity distance, and merger time. Unlike the simplified step-function memory model that tend to overestimate the final memory amplitude, our physically motivated waveforms allow more accurate signal recovery and parameter estimation.
We showed the effectiveness of our pipeline in recovering memory signals under various noise realizations and configurations through a series of simulation studies across realistic datasets. 
Even in the presence of degeneracies and multimodal posteriors, the Bayesian framework correctly captures these correlations and allows statistically consistent recovery of the injected source parameters. 
Our analysis provides a foundation for more detailed and physically motivated searches for SMBHB memory in PTA data.

\section*{Acknowledgements}

We thank Abhimanyu Sushobhanan and Wang Wei Yu for insightful scientific discussions. 
This work is supported by the Max Planck Gesellschaft (MPG) and the ATLAS cluster computing team at AEI Hannover.
\\
We made use of the following code.
To generate the gravitational-wave strain signal from SMBHB mergers including the null memory component, we use the publicly available \texttt{NRHybSur3dq8\_CCE} surrogate waveform model available as \texttt{gwsurrogate}~\citep{FieldVarma2025}. For simulating pulsar timing datasets with realistic observation spans, noise properties, and cadence, we use the \texttt{pta\_replicator} at \href{https://github.com/bencebecsy/pta_replicator.git}{github.com/bencebecsy/pta\_replicator} with an extension to include our SMBHB merger with memory waveform model. A tutorial and example scripts for simulating PTA datasets and recovering SMBHB merger signals with the waveform model used in this work are available at \href{https://github.com/sharontomson/SMBHB_Merger_Memory}{github.com/sharontomson/SMBHB\_Merger\_Memory}.
The Bayesian analysis is performed using the enterprise framework~\citep{EllisVallisneri2020}, the core data analysis software for pulsar timing arrays for computing likelihoods and posteriors. We build the custom signal model representing the full SMBHB waveform within this framework. We employ \texttt{enterprise\_warp} at \href{https://github.com/bvgoncharov/enterprise_warp}{github.com/bvgoncharov/enterprise\_warp} which is a wrapper interface that facilitates streamlined integration between \texttt{enterprise} \href{https://github.com/nanograv/enterprise} {(github.com/nanograv/enterprise)}, \texttt{libstempo}, \texttt{Tempo2}, \texttt{enterprise\_extensions} \href{https://github.com/nanograv/enterprise_extensions} {(github.com/nanograv/enterprise\_extensions)} and MCMC samplers. Posterior sampling for Bayesian inference is performed using the \texttt{PTMCMCSampler} package~\citep{EllisvanHaasteren2019}, which implements an efficient parallel tempering MCMC algorithm.

\section*{Data Availability}
The simulated PTA datasets that support the findings of this article are openly available in Zenodo ~\cite{tomson_2026_zenodo_pta}.

\bibliography{mybib,collab}

@ARTICLE{EPTA_CRN,
       author = {{Chen}, S. and {Caballero}, R.~N. and {Guo}, Y.~J. and {Chalumeau}, A. and {Liu}, K. and {Shaifullah}, G. and {Lee}, K.~J. and {Babak}, S. and {Desvignes}, G. and {Parthasarathy}, A. and {Hu}, H. and {van der Wateren}, E. and {Antoniadis}, J. and {Bak Nielsen}, A. -S. and {Bassa}, C.~G. and {Berthereau}, A. and {Burgay}, M. and {Champion}, D.~J. and {Cognard}, I. and {Falxa}, M. and {Ferdman}, R.~D. and {Freire}, P.~C.~C. and {Gair}, J.~R. and {Graikou}, E. and {Guillemot}, L. and {Jang}, J. and {Janssen}, G.~H. and {Karuppusamy}, R. and {Keith}, M.~J. and {Kramer}, M. and {Liu}, X.~J. and {Lyne}, A.~G. and {Main}, R.~A. and {McKee}, J.~W. and {Mickaliger}, M.~B. and {Perera}, B.~B.~P. and {Perrodin}, D. and {Petiteau}, A. and {Porayko}, N.~K. and {Possenti}, A. and {Samajdar}, A. and {Sanidas}, S.~A. and {Sesana}, A. and {Speri}, L. and {Stappers}, B.~W. and {Theureau}, G. and {Tiburzi}, C. and {Vecchio}, A. and {Verbiest}, J.~P.~W. and {Wang}, J. and {Wang}, L. and {Xu}, H.},
        title = "{Common-red-signal analysis with 24-yr high-precision timing of the European Pulsar Timing Array: inferences in the stochastic gravitational-wave background search}",
      journal = {\mnras},
         year = 2021,
        month = dec,
       volume = {508},
       number = {4},
        pages = {4970-4993},
          doi = {10.1093/mnras/stab2833},
archivePrefix = {arXiv},
       eprint = {2110.13184},
 primaryClass = {astro-ph.HE},
       adsurl = {https://ui.adsabs.harvard.edu/abs/2021MNRAS.508.4970C}
}

@ARTICLE{EPTA_DR2_GWB,
       author = {{EPTA Collaboration} and {InPTA Collaboration} and {Antoniadis}, J. and {Arumugam}, P. and {Arumugam}, S. and {Babak}, S. and {Bagchi}, M. and {Bak Nielsen}, A. -S. and {Bassa}, C.~G. and {Bathula}, A. and {Berthereau}, A. and {Bonetti}, M. and {Bortolas}, E. and {Brook}, P.~R. and {Burgay}, M. and {Caballero}, R.~N. and {Chalumeau}, A. and {Champion}, D.~J. and {Chanlaridis}, S. and {Chen}, S. and {Cognard}, I. and {Dandapat}, S. and {Deb}, D. and {Desai}, S. and {Desvignes}, G. and {Dhanda-Batra}, N. and {Dwivedi}, C. and {Falxa}, M. and {Ferdman}, R.~D. and {Franchini}, A. and {Gair}, J.~R. and {Goncharov}, \textit{et al.}},
        title = "{The second data release from the European Pulsar Timing Array. III. Search for gravitational wave signals}",
      journal = {\aap},
         year = 2023,
        month = oct,
       volume = {678},
          eid = {A50},
        pages = {A50},
          doi = {10.1051/0004-6361/202346844},
archivePrefix = {arXiv},
       eprint = {2306.16214},
 primaryClass = {astro-ph.HE},
       adsurl = {https://ui.adsabs.harvard.edu/abs/2023A\&A...678A..50E}
}

@ARTICLE{MT_DR1_GWB,
       author = {{Miles}, Matthew T. and {Shannon}, Ryan M. and {Reardon}, Daniel J. and {Bailes}, Matthew and {Champion}, David J. and {Geyer}, Marisa and {Gitika}, Pratyasha and {Grunthal}, Kathrin and {Keith}, Michael J. and {Kramer}, Michael and {Kulkarni}, Atharva D. and {Nathan}, Rowina S. and {Parthasarathy}, Aditya and {Singha}, Jaikhomba and {Theureau}, Gilles and {Thrane}, Eric and {Abbate}, Federico and {Buchner}, Sarah and {Cameron}, Andrew D. and {Camilo}, Fernando and {Moreschi}, Beatrice E. and {Shaifullah}, Golam and {Shamohammadi}, Mohsen and {Possenti}, Andrea and {Krishnan}, Vivek Venkatraman},
        title = "{The MeerKAT Pulsar Timing Array: the first search for gravitational waves with the MeerKAT radio telescope}",
      journal = {\mnras},
         year = 2025,
        month = jan,
       volume = {536},
       number = {2},
        pages = {1489-1500},
          doi = {10.1093/mnras/stae2571},
archivePrefix = {arXiv},
       eprint = {2412.01153},
 primaryClass = {astro-ph.HE},
       adsurl = {https://ui.adsabs.harvard.edu/abs/2025MNRAS.536.1489M}
}

@ARTICLE{PPTA_DR2_GWB,
       author = {{Goncharov}, Boris and {Shannon}, R.~M. and {Reardon}, D.~J. and {Hobbs}, G. and {Zic}, A. and {Bailes}, M. and {Cury{\l}o}, M. and {Dai}, S. and {Kerr}, M. and {Lower}, M.~E. and {Manchester}, R.~N. and {Mandow}, R. and {Middleton}, H. and {Miles}, M.~T. and {Parthasarathy}, A. and {Thrane}, E. and {Thyagarajan}, N. and {Xue}, X. and {Zhu}, X. -J. and {Cameron}, A.~D. and {Feng}, Y. and {Luo}, R. and {Russell}, C.~J. and {Sarkissian}, J. and {Spiewak}, R. and {Wang}, S. and {Wang}, J.~B. and {Zhang}, L. and {Zhang}, S.},
        title = "{On the Evidence for a Common-spectrum Process in the Search for the Nanohertz Gravitational-wave Background with the Parkes Pulsar Timing Array}",
      journal = {\apjl},
         year = 2021,
        month = aug,
       volume = {917},
       number = {2},
          eid = {L19},
        pages = {L19},
          doi = {10.3847/2041-8213/ac17f4},
archivePrefix = {arXiv},
       eprint = {2107.12112},
 primaryClass = {astro-ph.HE},
       adsurl = {https://ui.adsabs.harvard.edu/abs/2021ApJ...917L..19G}
}

@ARTICLE{PPTA_DR3_GWB,
       author = {{Reardon}, Daniel J. and {Zic}, Andrew and {Shannon}, Ryan M. and {Hobbs}, George B. and {Bailes}, Matthew and {Di Marco}, Valentina and {Kapur}, Agastya and {Rogers}, Axl F. and {Thrane}, Eric and {Askew}, Jacob and {Bhat}, N.~D. Ramesh and {Cameron}, Andrew and {Cury{\l}o}, Ma{\l}gorzata and {Coles}, William A. and {Dai}, Shi and {Goncharov}, Boris and {Kerr}, Matthew and {Kulkarni}, Atharva and {Levin}, Yuri and {Lower}, Marcus E. and {Manchester}, Richard N. and {Mandow}, Rami and {Miles}, Matthew T. and {Nathan}, Rowina S. and {Os{\l}owski}, Stefan and {Russell}, Christopher J. and {Spiewak}, Ren{\'e}e and {Zhang}, Songbo and {Zhu}, Xing-Jiang},
        title = "{Search for an Isotropic Gravitational-wave Background with the Parkes Pulsar Timing Array}",
      journal = {\apjl},
         year = 2023,
        month = jul,
       volume = {951},
       number = {1},
          eid = {L6},
        pages = {L6},
          doi = {10.3847/2041-8213/acdd02},
archivePrefix = {arXiv},
       eprint = {2306.16215},
 primaryClass = {astro-ph.HE},
       adsurl = {https://ui.adsabs.harvard.edu/abs/2023ApJ...951L...6R}
}

@ARTICLE{NG_15_GWB,
       author = {{Agazie}, Gabriella and {Anumarlapudi}, Akash and {Archibald}, Anne M. and {Arzoumanian}, Zaven and {Baker}, Paul T. and {B{\'e}csy}, Bence and {Blecha}, Laura and {Brazier}, Adam and {Brook}, Paul R. and {Burke-Spolaor}, Sarah and {Burnette}, Rand and {Case}, Robin and {Charisi}, Maria and {Chatterjee}, Shami and {Chatziioannou}, Katerina and {Cheeseboro}, Belinda D. and {Chen}, Siyuan and {Cohen}, Tyler and {Cordes}, James M. and {Cornish}, Neil J. and {Crawford}, Fronefield and {Cromartie}, H. Thankful and {Crowter}, Kathryn and {Cutler}, Curt J. and {Decesar}, Megan E. and {Degan}, Dallas and {Demorest}, Paul B. and {Deng}, Heling and {Dolch} \textit{et al.}},
        title = "{The NANOGrav 15 yr Data Set: Evidence for a Gravitational-wave Background}",
      journal = {\apjl},
         year = 2023,
        month = jul,
       volume = {951},
       number = {1},
          eid = {L8},
        pages = {L8},
          doi = {10.3847/2041-8213/acdac6},
archivePrefix = {arXiv},
       eprint = {2306.16213},
 primaryClass = {astro-ph.HE},
       adsurl = {https://ui.adsabs.harvard.edu/abs/2023ApJ...951L...8A}
}

@ARTICLE{NG_15_NEWPHYS,
       author = {{Afzal}, Adeela and {Agazie}, Gabriella and {Anumarlapudi}, Akash and {Archibald}, Anne M. and {Arzoumanian}, Zaven and {Baker}, Paul T. and {B{\'e}csy}, Bence and {Blanco-Pillado}, Jose Juan and {Blecha}, Laura and {Boddy}, Kimberly K. and {Brazier}, Adam and {Brook}, Paul R. and {Burke-Spolaor}, Sarah and {Burnette}, Rand and {Case}, Robin and {Charisi}, Maria and {Chatterjee}, Shami and {Chatziioannou}, Katerina and {Cheeseboro}, Belinda D. and {Chen}, Siyuan and {Cohen}, Tyler and {Cordes}, James M. and {Cornish}, Neil J. and {Crawford}, Fronefield and {Cromartie}, H. Thankful and {Crowter}, Kathryn and {Cutler}, Curt J. and {Decesar}, Megan E. and {Degan}, Dallas and {Demorest}, Paul B. and {Deng}, Heling \textit{et al.}},
        title = "{The NANOGrav 15 yr Data Set: Search for Signals from New Physics}",
      journal = {\apjl},
         year = 2023,
        month = jul,
       volume = {951},
       number = {1},
          eid = {L11},
        pages = {L11},
          doi = {10.3847/2041-8213/acdc91},
archivePrefix = {arXiv},
       eprint = {2306.16219},
 primaryClass = {astro-ph.HE},
       adsurl = {https://ui.adsabs.harvard.edu/abs/2023ApJ...951L..11A}
}

@ARTICLE{NG_15_MEM,
       author = {{Agazie}, Gabriella and {Anumarlapudi}, Akash and {Archibald}, Anne M. and {Arzoumanian}, Zaven and {Baier}, Jeremy G. and {Baker}, Paul T. and {B{\'e}csy}, Bence and {Blecha}, Laura and {Brazier}, Adam and {Brook}, Paul R. and {Burke-Spolaor}, Sarah and {Burnette}, Rand and {Casey-Clyde}, J. Andrew and {Charisi}, Maria and {Chatterjee}, Shami and {Cohen}, Tyler and {Cordes}, James M. and {Cornish}, Neil J. and {Crawford}, Fronefield and {Cromartie}, H. Thankful and {Crowter}, Kathryn and {DeCesar}, Megan E. and {Demorest}, Paul B. and {Deng}, Heling and {Dey}, Lankeswar and {Dolch}, Timothy and {Ferrara}, Elizabeth C. and {Fiore}, William and {Fonseca}, Emmanuel and {Freedman}, Gabriel E. and {Gardiner}, Emiko C. and {Garver-Daniels}, Nate and {Gentile}, Peter A. and {Gersbach}, Kyle A. and {Glaser}, Joseph and {Good}, Deborah C. and {G{\"u}ltekin}, Kayhan and {Hazboun}, Jeffrey S. \textit{et al.}},
        title = "{The NANOGrav 15 yr Data Set: Search for Gravitational-wave Memory}",
      journal = {\apj},
         year = 2025,
        month = jul,
       volume = {987},
       number = {1},
          eid = {5},
        pages = {5},
          doi = {10.3847/1538-4357/add874},
       adsurl = {https://ui.adsabs.harvard.edu/abs/2025ApJ...987....5A}
}

@ARTICLE{NG_12_GWB,
       author = {{Arzoumanian}, Zaven and {Baker}, Paul T. and {Blumer}, Harsha and {B{\'e}csy}, Bence and {Brazier}, Adam and {Brook}, Paul R. and {Burke-Spolaor}, Sarah and {Chatterjee}, Shami and {Chen}, Siyuan and {Cordes}, James M. and {Cornish}, Neil J. and {Crawford}, Fronefield and {Cromartie}, H. Thankful and {Decesar}, Megan E. and {Demorest}, Paul B. and {Dolch}, Timothy and {Ellis}, Justin A. and {Ferrara}, Elizabeth C. and {Fiore}, William \textit{et al.} and {Nanograv Collaboration}},
        title = "{The NANOGrav 12.5 yr Data Set: Search for an Isotropic Stochastic Gravitational-wave Background}",
      journal = {\apjl},
         year = 2020,
        month = dec,
       volume = {905},
       number = {2},
          eid = {L34},
        pages = {L34},
          doi = {10.3847/2041-8213/abd401},
archivePrefix = {arXiv},
       eprint = {2009.04496},
 primaryClass = {astro-ph.HE},
       adsurl = {https://ui.adsabs.harvard.edu/abs/2020ApJ...905L..34A}
}

@ARTICLE{NG_12_MEM,
       author = {{Agazie}, Gabriella and {Arzoumanian}, Zaven and {Baker}, Paul T. and {B{\'e}csy}, Bence and {Blecha}, Laura and {Blumer}, Harsha and {Brazier}, Adam and {Brook}, Paul R. and {Burke-Spolaor}, Sarah and {Burnette}, Rand and {Case}, Robin and {Casey-Clyde}, J. Andrew and {Charisi}, Maria and {Chatterjee}, Shami and {Cohen}, Tyler and {Cordes}, James M. and {Cornish}, Neil J. and {Crawford}, Fronefield and {Cromartie}, H. Thankful and {Decesar}, Megan E. and {Degan}, Dallas and {Demorest}, Paul B. and {Dolch}, Timothy and {Drachler}, Brendan and {Ellis}, Justin A. and {Ferdman}, Robert D. and {Ferrara}, Elizabeth C. and {Fiore}, William and {Fonseca}, Emmanuel and {Freedman}, Gabriel E. and {Garver-Daniels}, Nate and {Gentile}, Peter A. and {Glaser}, Joseph and {Good}, Deborah C. and {G{\"u}ltekin}, Kayhan and {Hazboun}, Jeffrey S. and {Jennings}, Ross J. \textit{et al.} and {The Nanograv Collaboration}},
        title = "{The NANOGrav 12.5 yr Data Set: Search for Gravitational Wave Memory}",
      journal = {\apj},
         year = 2024,
        month = mar,
       volume = {963},
       number = {1},
          eid = {61},
        pages = {61},
          doi = {10.3847/1538-4357/ad0726},
archivePrefix = {arXiv},
       eprint = {2502.18599},
 primaryClass = {gr-qc},
       adsurl = {https://ui.adsabs.harvard.edu/abs/2024ApJ...963...61A}
}

@article{NG_11_MEM,
   title={The NANOGrav 11 yr Data Set: Limits on Gravitational Wave Memory},
   volume={889},
   ISSN={1538-4357},
   url={http://dx.doi.org/10.3847/1538-4357/ab6083},
   DOI={10.3847/1538-4357/ab6083},
   number={1},
   journal={The Astrophysical Journal},
   publisher={American Astronomical Society},
   author={Aggarwal, K. and Arzoumanian, Z. and Baker, P. T. and Brazier, A. and Brook, P. R. and Burke-Spolaor, S. and Chatterjee, S. and Cordes, J. M. and Cornish, N. J. and Crawford, F. and Cromartie, H. T. and Crowter, K. and DeCesar, M. and Demorest, P. B. and Dolch, T. and Ellis, J. A. and Ferdman, R. D. and Ferrara, E. C. and Fonseca, E. and Garver-Daniels, N. and Gentile, P. and Good, D. \textit{et al.}},
   year={2020},
   month=jan, pages={38} }

@ARTICLE{CPTA_DR1_GWB,
       author = {{Xu}, Heng and {Chen}, Siyuan and {Guo}, Yanjun and {Jiang}, Jinchen and {Wang}, Bojun and {Xu}, Jiangwei and {Xue}, Zihan and {Caballero}, R. Nicolas and {Yuan}, Jianping and {Xu}, Yonghua and {Wang}, Jingbo and {Hao}, Longfei and {Luo}, Jingtao and {Lee}, Kejia and {Han}, Jinlin and {Jiang}, Peng and {Shen}, Zhiqiang and {Wang}, Min and {Wang}, Na and {Xu}, Renxin and {Wu}, Xiangping and {Manchester}, Richard and {Qian}, Lei and {Guan}, Xin and {Huang}, Menglin and {Sun}, Chun and {Zhu}, Yan},
        title = "{Searching for the Nano-Hertz Stochastic Gravitational Wave Background with the Chinese Pulsar Timing Array Data Release I}",
      journal = {Research in Astronomy and Astrophysics},
         year = 2023,
        month = jul,
       volume = {23},
       number = {7},
          eid = {075024},
        pages = {075024},
          doi = {10.1088/1674-4527/acdfa5},
archivePrefix = {arXiv},
       eprint = {2306.16216},
 primaryClass = {astro-ph.HE},
       adsurl = {https://ui.adsabs.harvard.edu/abs/2023RAA....23g5024X}
}

@ARTICLE{IPTA_DR2_GWB,
       author = {{Antoniadis}, J. and {Arzoumanian}, Z. and {Babak}, S. and {Bailes}, M. and {Bak Nielsen}, A. -S. and {Baker}, P.~T. and {Bassa}, C.~G. and {B{\'e}csy}, B. and {Berthereau}, A. and {Bonetti}, M. and {Brazier}, A. and {Brook}, P.~R. and {Burgay}, M. and {Burke-Spolaor}, S. and {Caballero}, R.~N. and {Casey-Clyde}, J.~A. and {Chalumeau}, A. and {Champion}, D.~J. and {Charisi}, M. and {Chatterjee}, S. and {Chen}, S. and {Cognard}, I. and {Cordes}, J.~M. and {Cornish}, N.~J. and {Crawford}, F. and {Cromartie}, H.~T. and {Crowter}, K. and {Dai}, S. and {DeCesar}, M.~E. and {Demorest}, P.~B. and {Desvignes}, G. and {Dolch}, T. and {Drachler}, B. and {Falxa}, M. and {Ferrara}, E.~C. and {Fiore}, W. and {Fonseca}, E. and {Gair}, J.~R. and {Garver-Daniels}, N. and {Goncharov}, B. \textit{et al.}},
        title = "{The International Pulsar Timing Array second data release: Search for an isotropic gravitational wave background}",
      journal = {\mnras},
         year = 2022,
        month = mar,
       volume = {510},
       number = {4},
        pages = {4873-4887},
          doi = {10.1093/mnras/stab3418},
archivePrefix = {arXiv},
       eprint = {2201.03980},
 primaryClass = {astro-ph.HE},
       adsurl = {https://ui.adsabs.harvard.edu/abs/2022MNRAS.510.4873A}
}

@ARTICLE{EPTA_SGWB,
       author = {{EPTA Collaboration} and {InPTA Collaboration} and {Antoniadis}, J. and {Arumugam}, P. and {Arumugam}, S. and {Babak}, S. and {Bagchi}, M. and {Bak Nielsen}, A. -S. and {Bassa}, C.~G. and {Bathula}, A. and {Berthereau}, A. and {Bonetti}, M. and {Bortolas}, E. and {Brook}, P.~R. and {Burgay}, M. and {Caballero}, R.~N. and {Chalumeau}, A. and {Champion}, D.~J. and {Chanlaridis}, S. and {Chen}, S. and {Cognard}, I. and {Dandapat}, S. and {Deb}, D. and {Desai}, S. and {Desvignes}, G. and {Dhanda-Batra}, N. and {Dwivedi}, C. and {Falxa}, M. and {Ferdman}, R.~D. and {Franchini}, A. and {Gair}, J.~R. and {Goncharov}, B. and {Gopakumar}, A. and {Graikou}, E. and {Grie{\ss}meier}, J. -M. and {Gualandris}, A. and {Guillemot}, L. and {Guo}, Y.~J. and {Gupta}, Y. and {Hisano}, S. and {Hu}, H. and {Iraci}, F.  \textit{et al.}},
      journal = {\aap},
         year = 2024,
        month = may,
       volume = {685},
          eid = {A94},
        pages = {A94},
          doi = {10.1051/0004-6361/202347433},
archivePrefix = {arXiv},
       eprint = {2306.16227},
 primaryClass = {astro-ph.CO},
       adsurl = {https://ui.adsabs.harvard.edu/abs/2024A&A...685A..94E}
}

@misc{tomson_2026_zenodo_pta,
  author    = {Tomson, Sharon Mary},
  title     = {Simulated Pulsar Timing Array Data for SMBHB Merger Searches with Full Waveform Memory},
  year      = {2026},
  publisher = {Zenodo},
  doi       = {10.5281/zenodo.19385397},
  url       = {https://doi.org/10.5281/zenodo.19385397}
}

@misc{tomson_github_smbhb_memory,
  author       = {Tomson, Sharon Mary},
  title        = {SMBHB\_Merger\_Memory},
  year         = {2026},
  howpublished = {\url{https://github.com/sharontomson/SMBHB_Merger_Memory}},
  note         = {GitHub repository},
  url          = {https://github.com/sharontomson/SMBHB_Merger_Memory}
}

@ARTICLE{CardinalTremblayGoncharov2025,
       author = {{Cardinal Tremblay}, Jacob and {Goncharov}, Boris and {van Haasteren}, Rutger and {Bhat}, N.~D. Ramesh and {Chen}, Zu-Cheng and {Di Marco}, Valentina and {Iguchi}, Satoru and {Kapur}, Agastya and {Ling}, Wenhua and {Mandow}, Rami and {Mishra}, Saurav and {Reardon}, Daniel J. and {Shannon}, Ryan M. and {Sudou}, Hiroshi and {Wang}, Jingbo and {Zhao}, Shi-Yi and {Zhu}, Xing-Jiang and {Zic}, Andrew},
        title = "{A Multi-Messenger Search for the Supermassive Black Hole Binary in 3C 66B with the Parkes Pulsar Timing Array}",
      journal = {arXiv e-prints},
         year = 2025,
        month = aug,
          eid = {arXiv:2508.20007},
        pages = {arXiv:2508.20007},
          doi = {10.48550/arXiv.2508.20007},
archivePrefix = {arXiv},
       eprint = {2508.20007},
 primaryClass = {astro-ph.HE},
       adsurl = {https://ui.adsabs.harvard.edu/abs/2025arXiv250820007C}
}

@ARTICLE{FieldVarma2025,
       author = {{Field}, Scott and {Varma}, Vijay and {Blackman}, Jonathan and {Gadre}, Bhooshan and {Galley}, Chad and {Islam}, Tousif and {Mitman}, Keefe and {P{\"u}rrer}, Michael and {Ravichandran}, Adhrit and {Scheel}, Mark and {Stein}, Leo and {Yoo}, Jooheon},
        title = "{GWSurrogate: A Python package for gravitational wave surrogate models}",
      journal = {The Journal of Open Source Software},
         year = 2025,
        month = mar,
       volume = {10},
       number = {107},
          eid = {7073},
        pages = {7073},
          doi = {10.21105/joss.07073},
archivePrefix = {arXiv},
       eprint = {2504.08839},
 primaryClass = {astro-ph.IM},
       adsurl = {https://ui.adsabs.harvard.edu/abs/2025JOSS...10.7073F}
}

@ARTICLE{DandapatSusobhanan2024,
       author = {{Dandapat}, Subhajit and {Susobhanan}, Abhimanyu and {Dey}, Lankeswar and {Gopakumar}, A. and {Baker}, Paul T. and {Jetzer}, Philippe},
        title = "{Efficient prescription to search for linear gravitational wave memory from hyperbolic black hole encounters and its application to the NANOGrav 12.5-year dataset}",
      journal = {\prd},
         year = 2024,
        month = may,
       volume = {109},
       number = {10},
          eid = {103018},
        pages = {103018},
          doi = {10.1103/PhysRevD.109.103018},
archivePrefix = {arXiv},
       eprint = {2402.03472},
 primaryClass = {astro-ph.HE},
       adsurl = {https://ui.adsabs.harvard.edu/abs/2024PhRvD.109j3018D}
}

@ARTICLE{YooMitman2023,
       author = {{Yoo}, Jooheon and {Mitman}, Keefe and {Varma}, Vijay and {Boyle}, Michael and {Field}, Scott E. and {Deppe}, Nils and {H{\'e}bert}, Fran{\c{c}}ois and {Kidder}, Lawrence E. and {Moxon}, Jordan and {Pfeiffer}, Harald P. and {Scheel}, Mark A. and {Stein}, Leo C. and {Teukolsky}, Saul A. and {Throwe}, William and {Vu}, Nils L.},
        title = "{Numerical relativity surrogate model with memory effects and post-Newtonian hybridization}",
      journal = {\prd},
         year = 2023,
        month = sep,
       volume = {108},
       number = {6},
          eid = {064027},
        pages = {064027},
          doi = {10.1103/PhysRevD.108.064027},
archivePrefix = {arXiv},
       eprint = {2306.03148},
 primaryClass = {gr-qc},
       adsurl = {https://ui.adsabs.harvard.edu/abs/2023PhRvD.108f4027Y}
}

@ARTICLE{MilesShannon2023,
       author = {{Miles}, M.~T. and {Shannon}, R.~M. and {Bailes}, M. and {Reardon}, D.~J. and {Keith}, M.~J. and {Cameron}, A.~D. and {Parthasarathy}, A. and {Shamohammadi}, M. and {Spiewak}, R. and {van Straten}, W. and {Buchner}, S. and {Camilo}, F. and {Geyer}, M. and {Karastergiou}, A. and {Kramer}, M. and {Serylak}, M. and {Theureau}, G. and {Venkatraman Krishnan}, V.},
        title = "{The MeerKAT Pulsar Timing Array: first data release}",
      journal = {\mnras},
         year = 2023,
        month = mar,
       volume = {519},
       number = {3},
        pages = {3976-3991},
          doi = {10.1093/mnras/stac3644},
archivePrefix = {arXiv},
       eprint = {2212.04648},
 primaryClass = {astro-ph.HE},
       adsurl = {https://ui.adsabs.harvard.edu/abs/2023MNRAS.519.3976M}
}

@ARTICLE{BogdanovicMiller2022,
       author = {{Bogdanovi{\'c}}, Tamara and {Miller}, M. Coleman and {Blecha}, Laura},
        title = "{Electromagnetic counterparts to massive black-hole mergers}",
      journal = {Living Reviews in Relativity},
         year = 2022,
        month = dec,
       volume = {25},
       number = {1},
          eid = {3},
        pages = {3},
          doi = {10.1007/s41114-022-00037-8},
archivePrefix = {arXiv},
       eprint = {2109.03262},
 primaryClass = {astro-ph.HE},
       adsurl = {https://ui.adsabs.harvard.edu/abs/2022LRR....25....3B}
}

@ARTICLE{TarafdarNobleson2022,
       author = {{Tarafdar}, Pratik and {Nobleson}, K. and {Rana}, Prerna and {Singha}, Jaikhomba and {Krishnakumar}, M.~A. and {Joshi}, Bhal Chandra and {Paladi}, Avinash Kumar and {Kolhe}, Neel and {Batra}, Neelam Dhanda and {Agarwal}, Nikita and {Bathula}, Adarsh and {Dandapat}, Subhajit and {Desai}, Shantanu and {Dey}, Lankeswar and {Hisano}, Shinnosuke and {Ingale}, Prathamesh and {Kato}, Ryo and {Kharbanda}, Divyansh and {Kikunaga}, Tomonosuke and {Marmat}, Piyush and {Pandian}, B. Arul and {Prabu}, T. and {Srivastava}, Aman and {Surnis}, Mayuresh and {Susarla}, Sai Chaitanya and {Susobhanan}, Abhimanyu and {Takahashi}, Keitaro and {Arumugam}, P. and {Bagchi}, Manjari and {Banik}, Sarmistha and {De}, Kishalay and {Girgaonkar}, Raghav and {Gopakumar}, A. and {Gupta}, Yashwant and {Maan}, Yogesh and {Manoharan}, P.~K. and {Naidu}, Arun and {Pathak}, Dhruv},
        title = "{The Indian Pulsar Timing Array: First data release}",
      journal = {\pasa},
         year = 2022,
        month = oct,
       volume = {39},
          eid = {e053},
        pages = {e053},
          doi = {10.1017/pasa.2022.46},
archivePrefix = {arXiv},
       eprint = {2206.09289},
 primaryClass = {astro-ph.IM},
       adsurl = {https://ui.adsabs.harvard.edu/abs/2022PASA...39...53T}
}

@ARTICLE{GoncharovThrane2022,
       author = {{Goncharov}, Boris and {Thrane}, Eric and {Shannon}, Ryan M. and {Harms}, Jan and {Bhat}, N.~D. Ramesh and {Hobbs}, George and {Kerr}, Matthew and {Manchester}, Richard N. and {Reardon}, Daniel J. and {Russell}, Christopher J. and {Zhu}, Xing-Jiang and {Zic}, Andrew},
        title = "{Consistency of the Parkes Pulsar Timing Array Signal with a Nanohertz Gravitational-wave Background}",
      journal = {\apjl},
         year = 2022,
        month = jun,
       volume = {932},
       number = {2},
          eid = {L22},
        pages = {L22},
          doi = {10.3847/2041-8213/ac76bb},
archivePrefix = {arXiv},
       eprint = {2206.03766},
 primaryClass = {gr-qc},
       adsurl = {https://ui.adsabs.harvard.edu/abs/2022ApJ...932L..22G}
}

@ARTICLE{GutierrezCombi2022,
       author = {{Guti{\'e}rrez}, Eduardo M. and {Combi}, Luciano and {Noble}, Scott C. and {Campanelli}, Manuela and {Krolik}, Julian H. and {L{\'o}pez Armengol}, Federico and {Garc{\'\i}a}, Federico},
        title = "{Electromagnetic Signatures from Supermassive Binary Black Holes Approaching Merger}",
      journal = {\apj},
         year = 2022,
        month = apr,
       volume = {928},
       number = {2},
          eid = {137},
        pages = {137},
          doi = {10.3847/1538-4357/ac56de},
archivePrefix = {arXiv},
       eprint = {2112.09773},
 primaryClass = {astro-ph.HE},
       adsurl = {https://ui.adsabs.harvard.edu/abs/2022ApJ...928..137G}
}

@ARTICLE{AntoniadisArzoumanian2022,
       author = {{Antoniadis}, J. and {Arzoumanian}, Z. and {Babak}, S. and {Bailes}, M. and {Bak Nielsen}, A. -S. and {Baker}, P.~T. and {Bassa}, C.~G. and {B{\'e}csy}, B. and {Berthereau}, A. and {Bonetti}, M. and {Brazier}, A. and {Brook}, P.~R. and {Burgay}, M. and {Burke-Spolaor}, S. and {Caballero}, R.~N. and {Casey-Clyde}, J.~A. and {Chalumeau} \textit{et al.}},
        title = "{The International Pulsar Timing Array second data release: Search for an isotropic gravitational wave background}",
      journal = {\mnras},
         year = 2022,
        month = mar,
       volume = {510},
       number = {4},
        pages = {4873-4887},
          doi = {10.1093/mnras/stab3418},
archivePrefix = {arXiv},
       eprint = {2201.03980},
 primaryClass = {astro-ph.HE},
       adsurl = {https://ui.adsabs.harvard.edu/abs/2022MNRAS.510.4873A}
}

@software{EllisVallisneri2020,
       author = {{Ellis}, Justin A. and {Vallisneri}, Michele and {Taylor}, Stephen R. and {Baker}, Paul T.},
        title = "{ENTERPRISE: Enhanced Numerical Toolbox Enabling a Robust PulsaR Inference SuitE}",
         year = 2020,
        month = sep,
          eid = {10.5281/zenodo.4059815},
          doi = {10.5281/zenodo.4059815},
      version = {v3.0.0},
    publisher = {Zenodo},
       adsurl = {https://ui.adsabs.harvard.edu/abs/2020zndo...4059815E}
}

@software{EllisvanHaasteren2019,
       author = {{Ellis}, Justin and {van Haasteren}, Rutger},
        title = "{PTMCMCSampler: Parallel tempering MCMC sampler package written in Python}",
 howpublished = {Astrophysics Source Code Library, record ascl:1912.017},
         year = 2019,
        month = dec,
          eid = {ascl:1912.017},
       url = {https://ui.adsabs.harvard.edu/abs/2019ascl.soft12017E}
}

@ARTICLE{AbbottAbbott2017,
       author = {{Abbott}, B.~P. and {Abbott}, R. and {Abbott}, T.~D. and {Abernathy}, M.~R. and {Acernese}, F. and {Ackley}, K. and {Adams}, C. and {Adams}, T. and {Addesso}, P. and {Adhikari}, R.~X. and {Adya}, V.~B. and {Affeldt}, C. and {Agathos}, M. and {Agatsuma}, K. and {Aggarwal}, N. and {Aguiar}, O.~D. and {Aiello}, L. and {Ain}, A. and {Ajith}, P. and {Allen}, B. and {Allocca}, A. and {Altin}, P.~A. and {Anderson}, S.~B. and {Anderson}, W.~G. and {Arai}, K. and {Araya}, M.~C. and {Arceneaux}, C.~C. and {Areeda}, J.~S. and {Arnaud}, N. and {Arun}, K.~G. and {Ascenzi}, S. and {Ashton}, G. and {Ast}, M. and {Aston}, S.~M. and {Astone}, P. and {Aufmuth}, P. and {Aulbert}, C. and {Babak}, S. and {Bacon}, P. and {Bader}, M.~K.~M. and {Baldaccini}, F. and {Ballardin}, G. and {Ballmer}, S.~W. and {Barayoga}, J.~C. and {Barclay}, S.~E. \textit{et al.}},
        title = "{The basic physics of the binary black hole merger GW150914}",
      journal = {Annalen der Physik},
         year = 2017,
        month = jan,
       volume = {529},
       number = {1-2},
          eid = {1600209},
        pages = {1600209},
          doi = {10.1002/andp.201600209},
archivePrefix = {arXiv},
       eprint = {1608.01940},
 primaryClass = {gr-qc},
       adsurl = {https://ui.adsabs.harvard.edu/abs/2017AnP...52900209A}
}

@ARTICLE{McKeeJanssen2016,
       author = {{McKee}, J.~W. and {Janssen}, G.~H. and {Stappers}, B.~W. and {Lyne}, A.~G. and {Caballero}, R.~N. and {Lentati}, L. and {Desvignes}, G. and {Jessner}, A. and {Jordan}, C.~A. and {Karuppusamy}, R. and {Kramer}, M. and {Cognard}, I. and {Champion}, D.~J. and {Graikou}, E. and {Lazarus}, P. and {Os{\l}owski}, S. and {Perrodin}, D. and {Shaifullah}, G. and {Tiburzi}, C. and {Verbiest}, J.~P.~W.},
        title = "{A glitch in the millisecond pulsar J0613-0200}",
      journal = {\mnras},
         year = 2016,
        month = sep,
       volume = {461},
       number = {3},
        pages = {2809-2817},
          doi = {10.1093/mnras/stw1442},
archivePrefix = {arXiv},
       eprint = {1606.04098},
 primaryClass = {astro-ph.HE},
       adsurl = {https://ui.adsabs.harvard.edu/abs/2016MNRAS.461.2809M}
}

@ARTICLE{VerbiestLentati2016,
       author = {{Verbiest}, J.~P.~W. and {Lentati}, L. and {Hobbs}, G. and {van Haasteren}, R. and {Demorest}, P.~B. and {Janssen}, G.~H. and {Wang}, J. -B. and {Desvignes}, G. and {Caballero}, R.~N. and {Keith}, M.~J. and {Champion}, D.~J. and {Arzoumanian}, Z. and {Babak}, S. and {Bassa}, C.~G. and {Bhat}, N.~D.~R. and {Brazier}, A. and {Brem}, P. and {Burgay}, M. and {Burke-Spolaor}, S. and {Chamberlin}, S.~J. and {Chatterjee}, S. and {Christy}, B. and {Cognard}, I. and {Cordes}, J.~M. and {Dai}, S. and {Dolch}, T. and {Ellis}, J.~A. and {Ferdman}, R.~D. and {Fonseca}, E. and {Gair}, J.~R. and {Garver-Daniels}, N.~E. and {Gentile}, P. and {Gonzalez}, M.~E. and {Graikou}, E. \textit{et al.}},
        title = "{The International Pulsar Timing Array: First data release}",
      journal = {\mnras},
         year = 2016,
        month = may,
       volume = {458},
       number = {2},
        pages = {1267-1288},
          doi = {10.1093/mnras/stw347},
archivePrefix = {arXiv},
       eprint = {1602.03640},
 primaryClass = {astro-ph.IM},
       adsurl = {https://ui.adsabs.harvard.edu/abs/2016MNRAS.458.1267V}
}

@ARTICLE{DesvignesCaballero2016,
       author = {{Desvignes}, G. and {Caballero}, R.~N. and {Lentati}, L. and {Verbiest}, J.~P.~W. and {Champion}, D.~J. and {Stappers}, B.~W. and {Janssen}, G.~H. and {Lazarus}, P. and {Os{\l}owski}, S. and {Babak}, S. and {Bassa}, C.~G. and {Brem}, P. and {Burgay}, M. and {Cognard}, I. and {Gair}, J.~R. and {Graikou}, E. and {Guillemot}, L. and {Hessels}, J.~W.~T. and {Jessner}, A. and {Jordan}, C. and {Karuppusamy}, R. and {Kramer}, M. and {Lassus}, A. and {Lazaridis}, K. and {Lee}, K.~J. and {Liu}, K. and {Lyne}, A.~G. and {McKee}, J. and {Mingarelli}, C.~M.~F. and {Perrodin}, D. and {Petiteau}, A. and {Possenti}, A. and {Purver}, M.~B. and {Rosado}, P.~A. and {Sanidas}, S. and {Sesana}, A. and {Shaifullah}, G. and {Smits}, R. and {Taylor}, S.~R. and {Theureau}, G. and {Tiburzi}, C. and {van Haasteren}, R. and {Vecchio}, A.},
        title = "{High-precision timing of 42 millisecond pulsars with the European Pulsar Timing Array}",
      journal = {\mnras},
         year = 2016,
        month = may,
       volume = {458},
       number = {3},
        pages = {3341-3380},
          doi = {10.1093/mnras/stw483},
archivePrefix = {arXiv},
       eprint = {1602.08511},
 primaryClass = {astro-ph.HE},
       adsurl = {https://ui.adsabs.harvard.edu/abs/2016MNRAS.458.3341D}
}

@ARTICLE{ArzoumanianBrazier2016,
       author = {{Arzoumanian}, Z. and {Brazier}, A. and {Burke-Spolaor}, S. and {Chamberlin}, S.~J. and {Chatterjee}, S. and {Christy}, B. and {Cordes}, J.~M. and {Cornish}, N.~J. and {Crowter}, K. and {Demorest}, P.~B. and {Deng}, X. and {Dolch}, T. and {Ellis}, J.~A. and {Ferdman}, R.~D. and {Fonseca}, E. and {Garver-Daniels}, N. and {Gonzalez}, M.~E. and {Jenet}, F. and {Jones}, G. and {Jones}, M.~L. and {Kaspi}, V.~M. and {Koop}, M. and {Lam}, M.~T. and {Lazio}, T.~J.~W. and {Levin}, L. and {Lommen}, A.~N. and {Lorimer}, D.~R. and {Luo}, J. and {Lynch}, R.~S. and {Madison}, D.~R. and {McLaughlin}, M.~A. and {McWilliams}, S.~T. and {Mingarelli}, C.~M.~F. and {Nice}, D.~J. and {Palliyaguru}, N. and {Pennucci}, T.~T. and {Ransom}, S.~M. and {Sampson}, L. and {Sanidas}, S.~A. and {Sesana}, A. and {Siemens}, X. and {Simon}, J. and {Stairs}, I.~H. and {Stinebring}, D.~R. and {Stovall}, K. and {Swiggum}, J. and {Taylor}, S.~R. and {Vallisneri}, M. and {van Haasteren}, R. and {Wang}, Y. and {Zhu}, W.~W. and {NANOGrav Collaboration}},
        title = "{The NANOGrav Nine-year Data Set: Limits on the Isotropic Stochastic Gravitational Wave Background}",
      journal = {\apj},
         year = 2016,
        month = apr,
       volume = {821},
       number = {1},
          eid = {13},
        pages = {13},
          doi = {10.3847/0004-637X/821/1/13},
archivePrefix = {arXiv},
       eprint = {1508.03024},
 primaryClass = {astro-ph.GA},
       adsurl = {https://ui.adsabs.harvard.edu/abs/2016ApJ...821...13A}
}

@INPROCEEDINGS{Lee2016,
       author = {{Lee}, K.~J.},
        title = "{Prospects of Gravitational Wave Detection Using Pulsar Timing Array for Chinese Future Telescopes}",
    booktitle = {Frontiers in Radio Astronomy and FAST Early Sciences Symposium 2015},
         year = 2016,
       editor = {{Qain}, L. and {Li}, D.},
       series = {Astronomical Society of the Pacific Conference Series},
       volume = {502},
        month = feb,
        pages = {19},
       adsurl = {https://ui.adsabs.harvard.edu/abs/2016ASPC..502...19L}
}

@ARTICLE{WangHobbs2015,
       author = {{Wang}, J.~B. and {Hobbs}, G. and {Coles}, W. and {Shannon}, R.~M. and {Zhu}, X.~J. and {Madison}, D.~R. and {Kerr}, M. and {Ravi}, V. and {Keith}, M.~J. and {Manchester}, R.~N. and {Levin}, Y. and {Bailes}, M. and {Bhat}, N.~D.~R. and {Burke-Spolaor}, S. and {Dai}, S. and {Os{\l}owski}, S. and {van Straten}, W. and {Toomey}, L. and {Wang}, N. and {Wen}, L.},
        title = "{Searching for gravitational wave memory bursts with the Parkes Pulsar Timing Array}",
      journal = {\mnras},
         year = 2015,
        month = jan,
       volume = {446},
       number = {2},
        pages = {1657-1671},
          doi = {10.1093/mnras/stu2137},
archivePrefix = {arXiv},
       eprint = {1410.3323},
 primaryClass = {astro-ph.GA},
       adsurl = {https://ui.adsabs.harvard.edu/abs/2015MNRAS.446.1657W}
}

@ARTICLE{McLaughlin2013,
       author = {{McLaughlin}, M.~A.},
        title = "{The North American Nanohertz Observatory for Gravitational Waves}",
      journal = {Classical and Quantum Gravity},
         year = 2013,
        month = nov,
       volume = {30},
       number = {22},
          eid = {224008},
        pages = {224008},
          doi = {10.1088/0264-9381/30/22/224008},
archivePrefix = {arXiv},
       eprint = {1310.0758},
 primaryClass = {astro-ph.IM},
       adsurl = {https://ui.adsabs.harvard.edu/abs/2013CQGra..30v4008M}
}

@ARTICLE{PetiteauBabak2013,
       author = {{Petiteau}, Antoine and {Babak}, Stanislav and {Sesana}, Alberto and {de Ara{\'u}jo}, Mariana},
        title = "{Resolving multiple supermassive black hole binaries with pulsar timing arrays. II. Genetic algorithm implementation}",
      journal = {\prd},
         year = 2013,
        month = mar,
       volume = {87},
       number = {6},
          eid = {064036},
        pages = {064036},
          doi = {10.1103/PhysRevD.87.064036},
archivePrefix = {arXiv},
       eprint = {1210.2396},
 primaryClass = {astro-ph.CO},
       adsurl = {https://ui.adsabs.harvard.edu/abs/2013PhRvD..87f4036P}
}

@ARTICLE{ManchesterHobbs2013,
       author = {{Manchester}, R.~N. and {Hobbs}, G. and {Bailes}, M. and {Coles}, W.~A. and {van Straten}, W. and {Keith}, M.~J. and {Shannon}, R.~M. and {Bhat}, N.~D.~R. and {Brown}, A. and {Burke-Spolaor}, S.~G. and {Champion}, D.~J. and {Chaudhary}, A. and {Edwards}, R.~T. and {Hampson}, G. and {Hotan}, A.~W. and {Jameson}, A. and {Jenet}, F.~A. and {Kesteven}, M.~J. and {Khoo}, J. and {Kocz}, J. and {Maciesiak}, K. and {Oslowski}, S. and {Ravi}, V. and {Reynolds}, J.~R. and {Sarkissian}, J.~M. and {Verbiest}, J.~P.~W. and {Wen}, Z.~L. and {Wilson}, W.~E. and {Yardley}, D. and {Yan}, W.~M. and {You}, X.~P.},
        title = "{The Parkes Pulsar Timing Array Project}",
      journal = {\pasa},
         year = 2013,
        month = jan,
       volume = {30},
          eid = {e017},
        pages = {e017},
          doi = {10.1017/pasa.2012.017},
archivePrefix = {arXiv},
       eprint = {1210.6130},
 primaryClass = {astro-ph.IM},
       adsurl = {https://ui.adsabs.harvard.edu/abs/2013PASA...30...17M}
}

@ARTICLE{EllisSiemens2012,
       author = {{Ellis}, J.~A. and {Siemens}, X. and {Creighton}, J.~D.~E.},
        title = "{Optimal Strategies for Continuous Gravitational Wave Detection in Pulsar Timing Arrays}",
      journal = {\apj},
         year = 2012,
        month = sep,
       volume = {756},
       number = {2},
          eid = {175},
        pages = {175},
          doi = {10.1088/0004-637X/756/2/175},
archivePrefix = {arXiv},
       eprint = {1204.4218},
 primaryClass = {astro-ph.IM},
       url = {https://ui.adsabs.harvard.edu/abs/2012ApJ...756..175E}
}

@ARTICLE{Favata2011,
       author = {{Favata}, Marc},
        title = "{The gravitational-wave memory from eccentric binaries}",
      journal = {\prd},
         year = 2011,
        month = dec,
       volume = {84},
       number = {12},
          eid = {124013},
        pages = {124013},
          doi = {10.1103/PhysRevD.84.124013},
archivePrefix = {arXiv},
       eprint = {1108.3121},
 primaryClass = {gr-qc},
       adsurl = {https://ui.adsabs.harvard.edu/abs/2011PhRvD..84l4013F}
}

@ARTICLE{PollneyReisswig2011,
       author = {{Pollney}, Denis and {Reisswig}, Christian},
        title = "{Gravitational Memory in Binary Black Hole Mergers}",
      journal = {\apjl},
         year = 2011,
        month = may,
       volume = {732},
       number = {1},
          eid = {L13},
        pages = {L13},
          doi = {10.1088/2041-8205/732/1/L13},
archivePrefix = {arXiv},
       eprint = {1004.4209},
 primaryClass = {gr-qc},
       adsurl = {https://ui.adsabs.harvard.edu/abs/2011ApJ...732L..13P}
}

@ARTICLE{Sesana2010,
       author = {{Sesana}, Alberto},
        title = "{Self Consistent Model for the Evolution of Eccentric Massive Black Hole Binaries in Stellar Environments: Implications for Gravitational Wave Observations}",
      journal = {\apj},
         year = 2010,
        month = aug,
       volume = {719},
       number = {1},
        pages = {851-864},
          doi = {10.1088/0004-637X/719/1/851},
archivePrefix = {arXiv},
       eprint = {1006.0730},
 primaryClass = {astro-ph.CO},
       adsurl = {https://ui.adsabs.harvard.edu/abs/2010ApJ...719..851S}
}

@ARTICLE{Favata2010,
       author = {{Favata}, Marc},
        title = "{The gravitational-wave memory effect}",
      journal = {Classical and Quantum Gravity},
         year = 2010,
        month = apr,
       volume = {27},
       number = {8},
          eid = {084036},
        pages = {084036},
          doi = {10.1088/0264-9381/27/8/084036},
archivePrefix = {arXiv},
       eprint = {1003.3486},
 primaryClass = {gr-qc},
       adsurl = {https://ui.adsabs.harvard.edu/abs/2010CQGra..27h4036F}
}

@ARTICLE{PshirkovBaskaran2010,
       author = {{Pshirkov}, M.~S. and {Baskaran}, D. and {Postnov}, K.~A.},
        title = "{Observing gravitational wave bursts in pulsar timing measurements}",
      journal = {\mnras},
         year = 2010,
        month = feb,
       volume = {402},
       number = {1},
        pages = {417-423},
          doi = {10.1111/j.1365-2966.2009.15887.x},
archivePrefix = {arXiv},
       eprint = {0909.0742},
 primaryClass = {astro-ph.CO},
       adsurl = {https://ui.adsabs.harvard.edu/abs/2010MNRAS.402..417P}
}

@ARTICLE{vanHaasterenLevin2010,
       author = {{van Haasteren}, Rutger and {Levin}, Yuri},
        title = "{Gravitational-wave memory and pulsar timing arrays}",
      journal = {\mnras},
         year = 2010,
        month = feb,
       volume = {401},
       number = {4},
        pages = {2372-2378},
          doi = {10.1111/j.1365-2966.2009.15885.x},
archivePrefix = {arXiv},
       eprint = {0909.0954},
 primaryClass = {astro-ph.IM},
       adsurl = {https://ui.adsabs.harvard.edu/abs/2010MNRAS.401.2372V}
}

@ARTICLE{Seto2009,
       author = {{Seto}, Naoki},
        title = "{Search for memory and inspiral gravitational waves from supermassive binary black holes with pulsar timing arrays}",
      journal = {\mnras},
         year = 2009,
        month = nov,
       volume = {400},
       number = {1},
        pages = {L38-L42},
          doi = {10.1111/j.1745-3933.2009.00758.x},
archivePrefix = {arXiv},
       eprint = {0909.1379},
 primaryClass = {astro-ph.CO},
       adsurl = {https://ui.adsabs.harvard.edu/abs/2009MNRAS.400L..38S}
}

@ARTICLE{vanHaasterenLevin2009,
       author = {{van Haasteren}, Rutger and {Levin}, Yuri and {McDonald}, Patrick and {Lu}, Tingting},
        title = "{On measuring the gravitational-wave background using Pulsar Timing Arrays}",
      journal = {\mnras},
         year = 2009,
        month = may,
       volume = {395},
       number = {2},
        pages = {1005-1014},
          doi = {10.1111/j.1365-2966.2009.14590.x},
archivePrefix = {arXiv},
       eprint = {0809.0791},
 primaryClass = {astro-ph},
       adsurl = {https://ui.adsabs.harvard.edu/abs/2009MNRAS.395.1005V}
}

@ARTICLE{Favata2009,
       author = {{Favata}, Marc},
        title = "{Nonlinear Gravitational-Wave Memory from Binary Black Hole Mergers}",
      journal = {\apjl},
         year = 2009,
        month = may,
       volume = {696},
       number = {2},
        pages = {L159-L162},
          doi = {10.1088/0004-637X/696/2/L159},
archivePrefix = {arXiv},
       eprint = {0902.3660},
 primaryClass = {astro-ph.SR},
       adsurl = {https://ui.adsabs.harvard.edu/abs/2009ApJ...696L.159F}
}

@ARTICLE{SagoIoka2004,
       author = {{Sago}, Norichika and {Ioka}, Kunihito and {Nakamura}, Takashi and {Yamazaki}, Ryo},
        title = "{Gravitational wave memory of gamma-ray burst jets}",
      journal = {\prd},
         year = 2004,
        month = nov,
       volume = {70},
       number = {10},
          eid = {104012},
        pages = {104012},
          doi = {10.1103/PhysRevD.70.104012},
archivePrefix = {arXiv},
       eprint = {gr-qc/0405067},
 primaryClass = {gr-qc},
       adsurl = {https://ui.adsabs.harvard.edu/abs/2004PhRvD..70j4012S}
}

@ARTICLE{DamourVilenkin2000,
       author = {{Damour}, Thibault and {Vilenkin}, Alexander},
        title = "{Gravitational Wave Bursts from Cosmic Strings}",
      journal = {\prl},
         year = 2000,
        month = oct,
       volume = {85},
       number = {18},
        pages = {3761-3764},
          doi = {10.1103/PhysRevLett.85.3761},
archivePrefix = {arXiv},
       eprint = {gr-qc/0004075},
 primaryClass = {gr-qc},
       adsurl = {https://ui.adsabs.harvard.edu/abs/2000PhRvL..85.3761D}
}

@ARTICLE{Christodoulou1991,
       author = {{Christodoulou}, Demetrios},
        title = "{Nonlinear nature of gravitation and gravitational-wave experiments}",
      journal = {\prl},
         year = 1991,
        month = sep,
       volume = {67},
        pages = {1486-1489},
          doi = {10.1103/PhysRevLett.67.1486},
       adsurl = {https://ui.adsabs.harvard.edu/abs/1991PhRvL..67.1486C}
}

@ARTICLE{FosterBacker1990,
       author = {{Foster}, R.~S. and {Backer}, D.~C.},
        title = "{Constructing a Pulsar Timing Array}",
      journal = {\apj},
         year = 1990,
        month = sep,
       volume = {361},
        pages = {300},
          doi = {10.1086/169195},
       adsurl = {https://ui.adsabs.harvard.edu/abs/1990ApJ...361..300F}
}

@ARTICLE{BraginskyThorne1987,
       author = {{Braginsky}, V.~B. and {Thorne}, Kip S.},
        title = "{Gravitational-wave bursts with memory and experimental prospects}",
      journal = {\nat},
         year = 1987,
        month = may,
       volume = {327},
       number = {6118},
        pages = {123-125},
          doi = {10.1038/327123a0},
       adsurl = {https://ui.adsabs.harvard.edu/abs/1987Natur.327..123B}
}

@ARTICLE{HellingsDowns1983,
       author = {{Hellings}, R.~W. and {Downs}, G.~S.},
        title = "{Upper limits on the isotropic gravitational radiation background from pulsar timing analysis.}",
      journal = {\apjl},
         year = 1983,
        month = feb,
       volume = {265},
        pages = {L39-L42},
          doi = {10.1086/183954},
       adsurl = {https://ui.adsabs.harvard.edu/abs/1983ApJ...265L..39H}
}

@ARTICLE{Epstein1978,
       author = {{Epstein}, R.},
        title = "{The generation of gravitational radiation by escaping supernova neutrinos.}",
      journal = {\apj},
         year = 1978,
        month = aug,
       volume = {223},
        pages = {1037-1045},
          doi = {10.1086/156337},
       adsurl = {https://ui.adsabs.harvard.edu/abs/1978ApJ...223.1037E}
}

@ARTICLE{Turner1977,
       author = {{Turner}, M.},
        title = "{Gravitational radiation from point-masses in unbound orbits: Newtonian results.}",
      journal = {\apj},
         year = 1977,
        month = sep,
       volume = {216},
        pages = {610-619},
          doi = {10.1086/155501},
       adsurl = {https://ui.adsabs.harvard.edu/abs/1977ApJ...216..610T}
}

@ARTICLE{BishopGomez1997,
       author = {{Bishop}, N. and {Gomez}, R. and {Lehner}, L. and {Winicour}, J.},
        title = "{Cauchy-characteristic extraction in numerical relativity}",
      journal = {arXiv e-prints},
         year = 1997,
        month = may,
          eid = {gr-qc/9705033},
        pages = {gr-qc/9705033},
          doi = {10.48550/arXiv.gr-qc/9705033},
archivePrefix = {arXiv},
       eprint = {gr-qc/9705033},
 primaryClass = {gr-qc},
       adsurl = {https://ui.adsabs.harvard.edu/abs/1997gr.qc.....5033B}
}

@ARTICLE{MoxonScheel2023,
       author = {{Moxon}, Jordan and {Scheel}, Mark A. and {Teukolsky}, Saul A. and {Deppe}, Nils and {Vu}, Nils and {H{\'e}bert}, Francois and {Kidder}, Lawrence E. and {Throwe}, William},
        title = "{SpECTRE Cauchy-characteristic evolution system for rapid, precise waveform extraction}",
      journal = {\prd},
         year = 2023,
        month = mar,
       volume = {107},
       number = {6},
          eid = {064013},
        pages = {064013},
          doi = {10.1103/PhysRevD.107.064013},
       adsurl = {https://ui.adsabs.harvard.edu/abs/2023PhRvD.107f4013M}
}

@ARTICLE{MitmanMoxon2020,
       author = {{Mitman}, Keefe and {Moxon}, Jordan and {Scheel}, Mark A. and {Teukolsky}, Saul A. and {Boyle}, Michael and {Deppe}, Nils and {Kidder}, Lawrence E. and {Throwe}, William},
        title = "{Computation of displacement and spin gravitational memory in numerical relativity}",
      journal = {\prd},
         year = 2020,
        month = nov,
       volume = {102},
       number = {10},
          eid = {104007},
        pages = {104007},
          doi = {10.1103/PhysRevD.102.104007},
archivePrefix = {arXiv},
       eprint = {2007.11562},
 primaryClass = {gr-qc},
       adsurl = {https://ui.adsabs.harvard.edu/abs/2020PhRvD.102j4007M}
}

@ARTICLE{ColesHobbs2011,
       author = {{Coles}, W. and {Hobbs}, G. and {Champion}, D.~J. and {Manchester}, R.~N. and {Verbiest}, J.~P.~W.},
        title = "{Pulsar timing analysis in the presence of correlated noise}",
      journal = {\mnras},
         year = 2011,
        month = nov,
       volume = {418},
       number = {1},
        pages = {561-570},
          doi = {10.1111/j.1365-2966.2011.19505.x},
archivePrefix = {arXiv},
       eprint = {1107.5366},
 primaryClass = {astro-ph.IM},
       adsurl = {https://ui.adsabs.harvard.edu/abs/2011MNRAS.418..561C}
}

@ARTICLE{ShannonCordes2010,
       author = {{Shannon}, Ryan M. and {Cordes}, James M.},
        title = "{Assessing the Role of Spin Noise in the Precision Timing of Millisecond Pulsars}",
      journal = {\apj},
         year = 2010,
        month = dec,
       volume = {725},
       number = {2},
        pages = {1607-1619},
          doi = {10.1088/0004-637X/725/2/1607},
archivePrefix = {arXiv},
       eprint = {1010.4794},
 primaryClass = {astro-ph.SR},
       adsurl = {https://ui.adsabs.harvard.edu/abs/2010ApJ...725.1607S}
}

@ARTICLE{LentatiAlexander2013,
       author = {{Lentati}, Lindley and {Alexander}, P. and {Hobson}, M.~P. and {Taylor}, S. and {Gair}, J. and {Balan}, S.~T. and {van Haasteren}, R.},
        title = "{Hyper-efficient model-independent Bayesian method for the analysis of pulsar timing data}",
      journal = {\prd},
         year = 2013,
        month = may,
       volume = {87},
       number = {10},
          eid = {104021},
        pages = {104021},
          doi = {10.1103/PhysRevD.87.104021},
archivePrefix = {arXiv},
       eprint = {1210.3578},
 primaryClass = {astro-ph.IM},
       adsurl = {https://ui.adsabs.harvard.edu/abs/2013PhRvD..87j4021L}
}

@ARTICLE{vanHaasterenVallisneri2015,
       author = {{van Haasteren}, Rutger and {Vallisneri}, Michele},
        title = "{Low-rank approximations for large stationary covariance matrices, as used in the Bayesian and generalized-least-squares analysis of pulsar-timing data}",
      journal = {\mnras},
         year = 2015,
        month = jan,
       volume = {446},
       number = {2},
        pages = {1170-1174},
          doi = {10.1093/mnras/stu2157},
archivePrefix = {arXiv},
       eprint = {1407.6710},
 primaryClass = {astro-ph.IM},
       adsurl = {https://ui.adsabs.harvard.edu/abs/2015MNRAS.446.1170V}
}

@ARTICLE{AllenRomano2025,
       author = {{Allen}, Bruce and {Romano}, Joseph D.},
        title = "{Optimal Reconstruction of the Hellings and Downs Correlation}",
      journal = {\prl},
         year = 2025,
        month = jan,
       volume = {134},
       number = {3},
          eid = {031401},
        pages = {031401},
          doi = {10.1103/PhysRevLett.134.031401},
archivePrefix = {arXiv},
       eprint = {2407.10968},
 primaryClass = {gr-qc},
       adsurl = {https://ui.adsabs.harvard.edu/abs/2025PhRvL.134c1401A}
}

@ARTICLE{CrisostomivanHaasteren2025,
       author = {{Crisostomi}, Marco and {van Haasteren}, Rutger and {Meyers}, Patrick M. and {Vallisneri}, Michele},
        title = "{Beyond diagonal approximations: improved covariance modeling for pulsar timing array data analysis}",
      journal = {arXiv e-prints},
         year = 2025,
        month = jun,
          eid = {arXiv:2506.13866},
        pages = {arXiv:2506.13866},
          doi = {10.48550/arXiv.2506.13866},
archivePrefix = {arXiv},
       eprint = {2506.13866},
 primaryClass = {astro-ph.IM},
       adsurl = {https://ui.adsabs.harvard.edu/abs/2025arXiv250613866C}
}

@ARTICLE{DandapatEbersold2023,
       author = {{Dandapat}, Subhajit and {Ebersold}, Michael and {Susobhanan}, Abhimanyu and {Rana}, Prerna and {Gopakumar}, A. and {Tiwari}, Shubhanshu and {Haney}, Maria and {Lee}, Hyung Mok and {Kolhe}, Neel},
        title = "{Gravitational waves from black-hole encounters: Prospects for ground and galaxy-based observatories}",
      journal = {\prd},
         year = 2023,
        month = jul,
       volume = {108},
       number = {2},
          eid = {024013},
        pages = {024013},
          doi = {10.1103/PhysRevD.108.024013},
archivePrefix = {arXiv},
       eprint = {2305.19318},
 primaryClass = {gr-qc},
       adsurl = {https://ui.adsabs.harvard.edu/abs/2023PhRvD.108b4013D}
}

@ARTICLE{MadisonChernoff2017,
       author = {{Madison}, Dustin R. and {Chernoff}, David F. and {Cordes}, James M.},
        title = "{Pulsar timing perturbations from Galactic gravitational wave bursts with memory}",
      journal = {\prd},
         year = 2017,
        month = dec,
       volume = {96},
       number = {12},
          eid = {123016},
        pages = {123016},
          doi = {10.1103/PhysRevD.96.123016},
archivePrefix = {arXiv},
       eprint = {1710.04974},
 primaryClass = {astro-ph.GA},
       adsurl = {https://ui.adsabs.harvard.edu/abs/2017PhRvD..96l3016M}
}

@ARTICLE{XiaWang2025,
       author = {{Xia}, Yong and {Wang}, Jingbo and {Kuroyanagi}, Sachiko and {Yan}, Wenming and {Wen}, Yirong and {Kapur}, Agastya and {Zou}, Jing and {Feng}, Yi and {Di Marco}, Valentina and {Mishra}, Saurav and {Russell}, Christopher J. and {Wang}, Shuangqiang and {Zhao}, De and {Zhu}, Xingjiang},
        title = "{Searching for Gravitational-Wave Bursts from Cosmic String Cusps with the Parkes Pulsar Timing Array's Third Data Release}",
      journal = {Universe},
         year = 2025,
        month = mar,
       volume = {11},
       number = {3},
          eid = {81},
        pages = {81},
          doi = {10.3390/universe11030081},
archivePrefix = {arXiv},
       eprint = {2502.21069},
 primaryClass = {gr-qc},
       adsurl = {https://ui.adsabs.harvard.edu/abs/2025Univ...11...81X}
}

@ARTICLE{TaylorBurnette2025,
       author = {{Taylor}, Jacob A. and {Burnette}, Rand and {B{\'e}csy}, Bence and {Cornish}, Neil J.},
        title = "{Fast wavelet basis search for generic gravitational wave bursts in pulsar timing array data}",
      journal = {\prd},
         year = 2025,
        month = jan,
       volume = {111},
       number = {2},
          eid = {022006},
        pages = {022006},
          doi = {10.1103/PhysRevD.111.022006},
archivePrefix = {arXiv},
       eprint = {2408.07864},
 primaryClass = {gr-qc},
       adsurl = {https://ui.adsabs.harvard.edu/abs/2025PhRvD.111b2006T}
}

@INPROCEEDINGS{EllisJenet2012,
       author = {{Ellis}, Justin and {Jenet}, F. and {Siemens}, X.},
        title = "{Detection Methods for Continuous Gravitational Waves using Pulsar Timing Data}",
    booktitle = {American Astronomical Society Meeting Abstracts \#219},
         year = 2012,
       series = {American Astronomical Society Meeting Abstracts},
       volume = {219},
        month = jan,
          eid = {146.20},
        pages = {146.20},
       adsurl = {https://ui.adsabs.harvard.edu/abs/2012AAS...21914620E}
}

@ARTICLE{MilesShannon2025,
       author = {{Miles}, Matthew T. and {Shannon}, Ryan M. and {Reardon}, Daniel J. and {Bailes}, Matthew and {Champion}, David J. and {Geyer}, Marisa and {Gitika}, Pratyasha and {Grunthal}, Kathrin and {Keith}, Michael J. and {Kramer}, Michael and {Kulkarni}, Atharva D. and {Nathan}, Rowina S. and {Parthasarathy}, Aditya and {Porayko}, Nataliya K. and {Singha}, Jaikhomba and {Theureau}, Gilles and {Abbate}, Federico and {Buchner}, Sarah and {Cameron}, Andrew D. and {Camilo}, Fernando and {Moreschi}, Beatrice E. and {Shaifullah}, Golam and {Shamohammadi}, Mohsen and {Krishnan}, Vivek Venkatraman},
        title = "{The MeerKAT Pulsar Timing Array: the 4.5-yr data release and the noise and stochastic signals of the millisecond pulsar population}",
      journal = {\mnras},
         year = 2025,
        month = jan,
       volume = {536},
       number = {2},
        pages = {1467-1488},
          doi = {10.1093/mnras/stae2572},
archivePrefix = {arXiv},
       eprint = {2412.01148},
 primaryClass = {astro-ph.HE},
       adsurl = {https://ui.adsabs.harvard.edu/abs/2025MNRAS.536.1467M}
}

@ARTICLE{AgarwalAgazie2025,
       author = {{Agarwal}, Nikita and {Agazie}, Gabriella and {Anumarlapudi}, Akash and {Archibald}, Anne M. and {Arzoumanian}, Zaven and {Baier}, Jeremy G. and {Baker}, Paul T. and {Becsy}, Bence and {Blecha}, Laura and {Brazier}, Adam and {Brook}, Paul R. and {Burke-Spolaor}, Sarah and {Burnette}, Rand and {Case}, Robin and {Casey-Clyde}, J. Andrew and {Chang}, Yu-Ting and {Charisi}, Maria and {Chatterjee}, Shami and {Cohen}, Tyler and {Coppi}, Paolo and {Cordes}, James M. and {Cornish}, Neil J. and {Crawford}, Fronefield and {Cromartie}, H. Thankful and {Crowter}, Kathryn and {DeCesar}, Megan E. and {Demorest}, Paul B. and {Deng}, Heling and {Dey}, Lankeswar and {Dolch}, Timothy and {D'Orazio}, Daniel J. and {Eisenberg}, Ellis and {Ferrara}, Elizabeth C. and {Fiore}, William and {Fonseca}, Emmanuel and {Freedman}, Gabriel E. and {Gardiner}, Emiko C. and {Garver-Daniels}, Nate and {Gentile}, Peter A. and {Gersbach}, Kyle A. and {Glaser}, Joseph and {Graham}, Matthew J. and {Good}, Deborah C. and {Gultekin}, Kayhan and {Harris}, C.~J. and {Hazboun}, Jeffrey S. and {Hutchison}, Forrest and {Jennings}, Ross J. and {Johnson}, Aaron D. and {Jones}, Megan L. and {Kaplan}, David L. and {Kelley}, Luke Zoltan and {Kerr}, Matthew and {Key}, Joey S. and {Laal}, Nima and {Lam}, Michael T. and {Lamb}, William G. and {Larsen}, Bjorn and {Lazio}, T. Joseph W. and {Lewandowska}, Natalia and {Liu}, Tingting and {Lorimer}, Duncan R. and {Luo}, Jing and {Lynch}, Ryan S. and {Ma}, Chung-Pei and {Madison}, Dustin R. and {Matt}, Cayenne and {McEwen}, Alexander and {McKee}, James W. and {McLaughlin}, Maura A. and {McMann}, Natasha and {Meyers}, Bradley W. and {Meyers}, Patrick M. and {Mingarelli}, Chiara M.~F. and {Mitridate}, Andrea and {Natarajan}, Priyamvada and {Ng}, Cherry and {Nice}, David J. and {Ocker}, Stella Koch and {Olum}, Ken D. and {Pennucci}, Timothy T. and {Perera}, Benetge B.~P. and {Petrov}, Polina and {Pol}, Nihan S. and {Radovan}, Henri A. and {Ransom}, Scott M. and {Ray}, Paul S. and {Romano}, Joseph D. and {Runnoe}, Jessie C. and {Saffer}, Alexander and {Sardesai}, Shashwat C. and {Schmiedekamp}, Ann and {Schmiedekamp}, Carl and {Schmitz}, Kai and {Semenzato}, Federico and {Shapiro-Albert}, Brent J. and {Shivakumar}, Rohan and {Siemens}, Xavier and {Simon}, Joseph and {Sosa Fiscella}, Sophia V. and {Stairs}, Ingrid H. and {Stinebring}, Daniel R. and {Stovall}, Kevin and {Susobhanan}, Abhimanyu and {Swiggum}, Joseph K. and {Taylor}, Jacob A. and {Taylor}, Stephen R. and {Thompson}, Mercedes S. and {Turner}, Jacob E. and {Vallisneri}, Michele and {van Haasteren}, Rutger and {Vigeland}, Sarah J. and {Wahl}, Haley M. and {Willson}, London and {Wilson}, Kevin P. and {Witt}, Caitlin A. and {Wright}, David and {Young}, Olivia and {Zheng}, Qinyuan},
        title = "{The NANOGrav 15 yr Data Set: Targeted Searches for Supermassive Black Hole Binaries}",
      journal = {arXiv e-prints},
         year = 2025,
        month = aug,
          eid = {arXiv:2508.16534},
        pages = {arXiv:2508.16534},
          doi = {10.48550/arXiv.2508.16534},
archivePrefix = {arXiv},
       eprint = {2508.16534},
 primaryClass = {astro-ph.HE},
       adsurl = {https://ui.adsabs.harvard.edu/abs/2025arXiv250816534A}
}

@ARTICLE{TomsonGoncharov2026,
       author = {{Tomson}, Sharon Mary and {Goncharov}, Boris and {van Haasteren}, Rutger and {Srinivasan}, Rahul and {Barausse}, Enrico and {Wen}, Yirong and {Wang}, Jingbo and {Antoniadis}, John and {Bhat}, N.~D. Ramesh and {Chen}, Zu-Cheng and {Cognard}, Ismael and {Di Marco}, Valentina and {Hu}, Huanchen and {Janssen}, Gemma H. and {Kramer}, Michael and {Ling}, Wenhua and {Liu}, Kuo and {Mishra}, Saurav and {Perrodin}, Delphine and {Possenti}, Andrea and {Russell}, Christopher J. and {Shannon}, Ryan M. and {Theureau}, Gilles and {Wang}, Shuangqiang},
        title = "{Search for Gravitational-wave Memory in PPTA and EPTA Data: A Complete Signal Model}",
      journal = {\apjl},
         year = 2026,
        month = jan,
       volume = {996},
       number = {1},
          eid = {L9},
        pages = {L9},
          doi = {10.3847/2041-8213/ae2912},
archivePrefix = {arXiv},
       eprint = {2512.14650},
 primaryClass = {gr-qc},
       adsurl = {https://ui.adsabs.harvard.edu/abs/2026ApJ...996L...9T}
}

@misc{enterprise,
  author       = {Stephen R. Taylor and Paul T. Baker and Jeffrey S. Hazboun and Joseph Simon and Sarah J. Vigeland},
  title        = {enterprise\_extensions},
  year         = {2021},
  url          = {https://github.com/nanograv/enterprise_extensions},
  note         = {v2.4.3}
}

@ARTICLE{HeeHandley2016,
       author = {{Hee}, S. and {Handley}, W.~J. and {Hobson}, M.~P. and {Lasenby}, A.~N.},
        title = "{Bayesian model selection without evidences: application to the dark energy equation-of-state}",
      journal = {\mnras},
         year = 2016,
        month = jan,
       volume = {455},
       number = {3},
        pages = {2461-2473},
          doi = {10.1093/mnras/stv2217},
archivePrefix = {arXiv},
       eprint = {1506.09024},
 primaryClass = {astro-ph.CO},
       adsurl = {https://ui.adsabs.harvard.edu/abs/2016MNRAS.455.2461H}
}

\clearpage 
\onecolumngrid
\appendix

\section{Appendix}

\subsection{Posterior distributions}

The appendix presents the full posterior distributions for all simulation runs discussed in the main text, including the recovered sky locations, masses, distances, and other model parameters. These results provide a complete view of the parameter estimation performance of our SMBHB merger model across different simulated scenarios.

\begin{figure*}[htbp]
  \centering
  \includegraphics[width=0.95\textwidth]{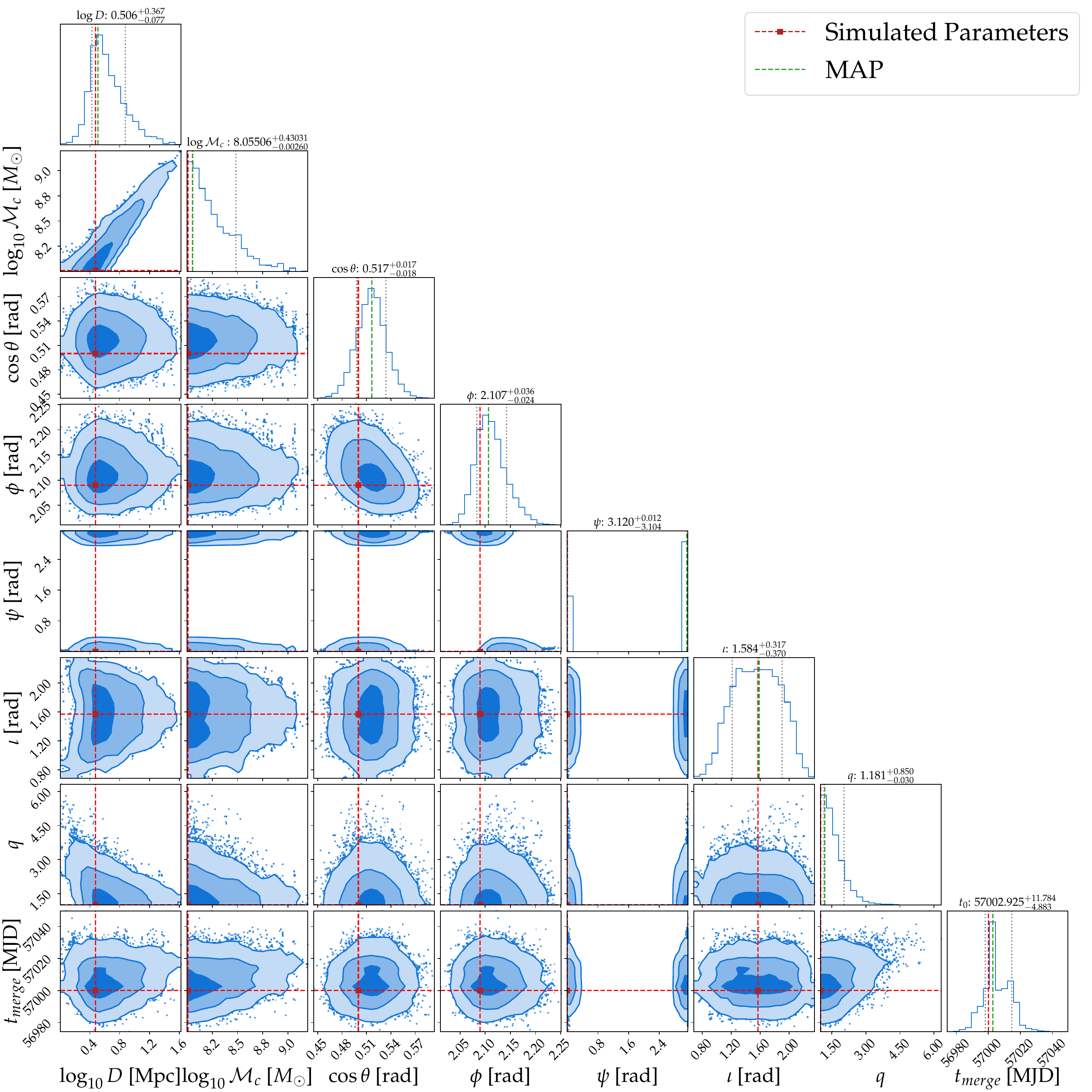}  
  \caption{\justifying
 Posterior distributions of SMBHB merger with null memory parameters from simulation studies on 25 pulsars. Parameters of the binary used for simulation are $\mathcal{M}_c = 10^{8} M_\odot$ and $D_L = 3$ Mpc. Posteriors shown in blue is the recovery of the signal from the \texttt{wn+mem+rn} dataset. There are clear covariances between parameters especially the chirp mass and the mass ratio of the binary. Red lines are the simulated values of the SMBHB merger and green lines are the maximum-\textit{a-posteriori} (MAP) estimates from the recovered posterior. The grey dotted lines on the diagonal panels denote the 68\% credible intervals (1$\sigma$) for each parameter.
  }
  \label{fig:sim_wn_rn_m8}
\end{figure*}

\begin{figure*}[htbp]
  \centering
  \includegraphics[width=0.97\textwidth]{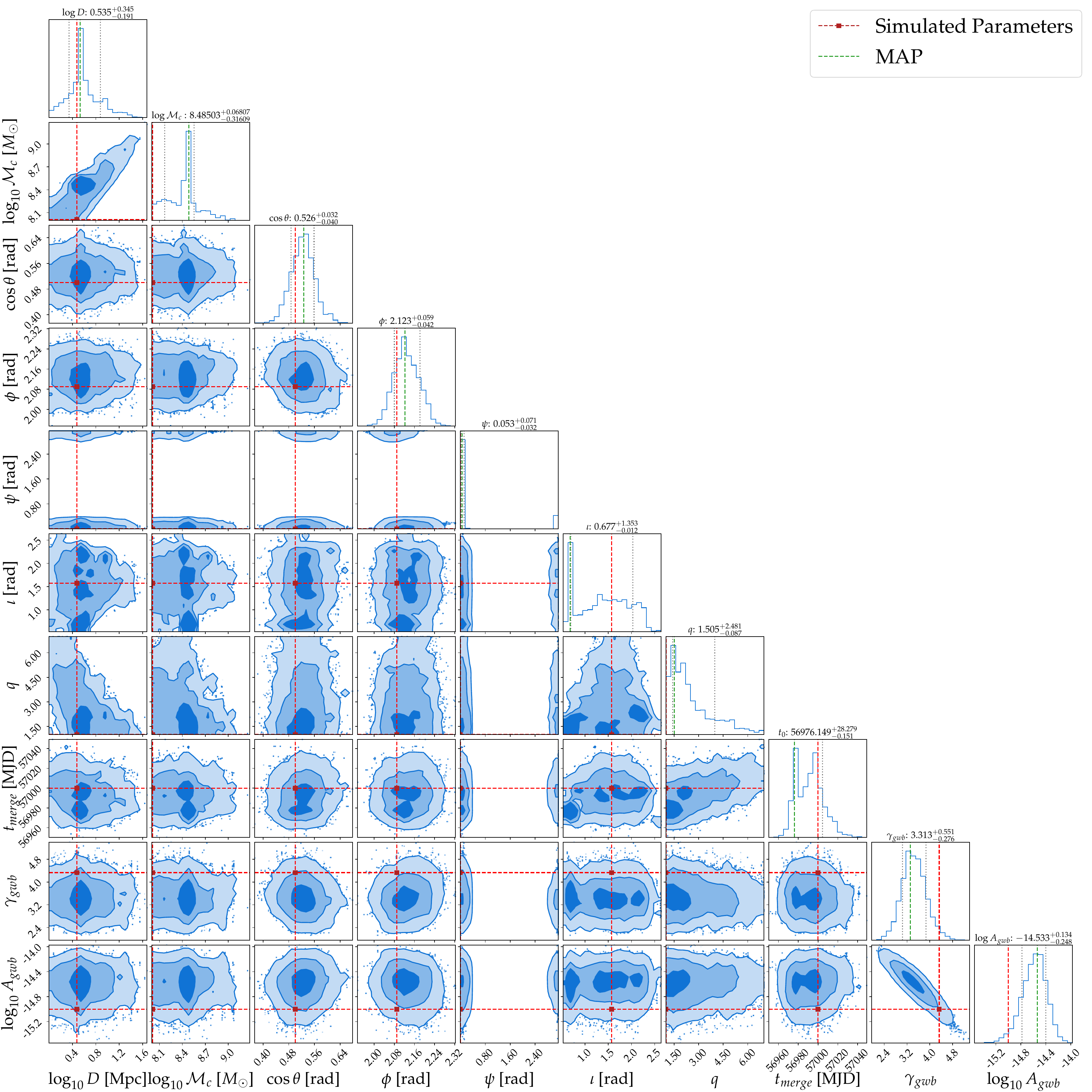}  
  \caption{\justifying
 Posterior distribution of the SMBHB merger from simulation studies on 25 pulsars. The dataset used is \texttt{wn+mem+rn+gwb} as described in this paper. Parameters of the binary used for simulation are $\mathcal{M}_c = 10^{8} M_\odot$ and $D_L = 3$ Mpc. Overplotted red lines indicate the simulated values used in the simulated data, and green lines mark the maximum-\textit{a-posteriori} (MAP) estimates from the recovered posterior. Grey dotted lines on the denote the 68\% credible intervals (1$\sigma$) for each parameter. All recovered parameters agree with the simulated values within a percentage uncertainty of $\lesssim 0.1\%$.
  }
  \label{fig:sim_wn_rn_gwb_m8}
\end{figure*}

\begin{figure*}[htbp]
  \centering
  \includegraphics[width=0.97\textwidth]{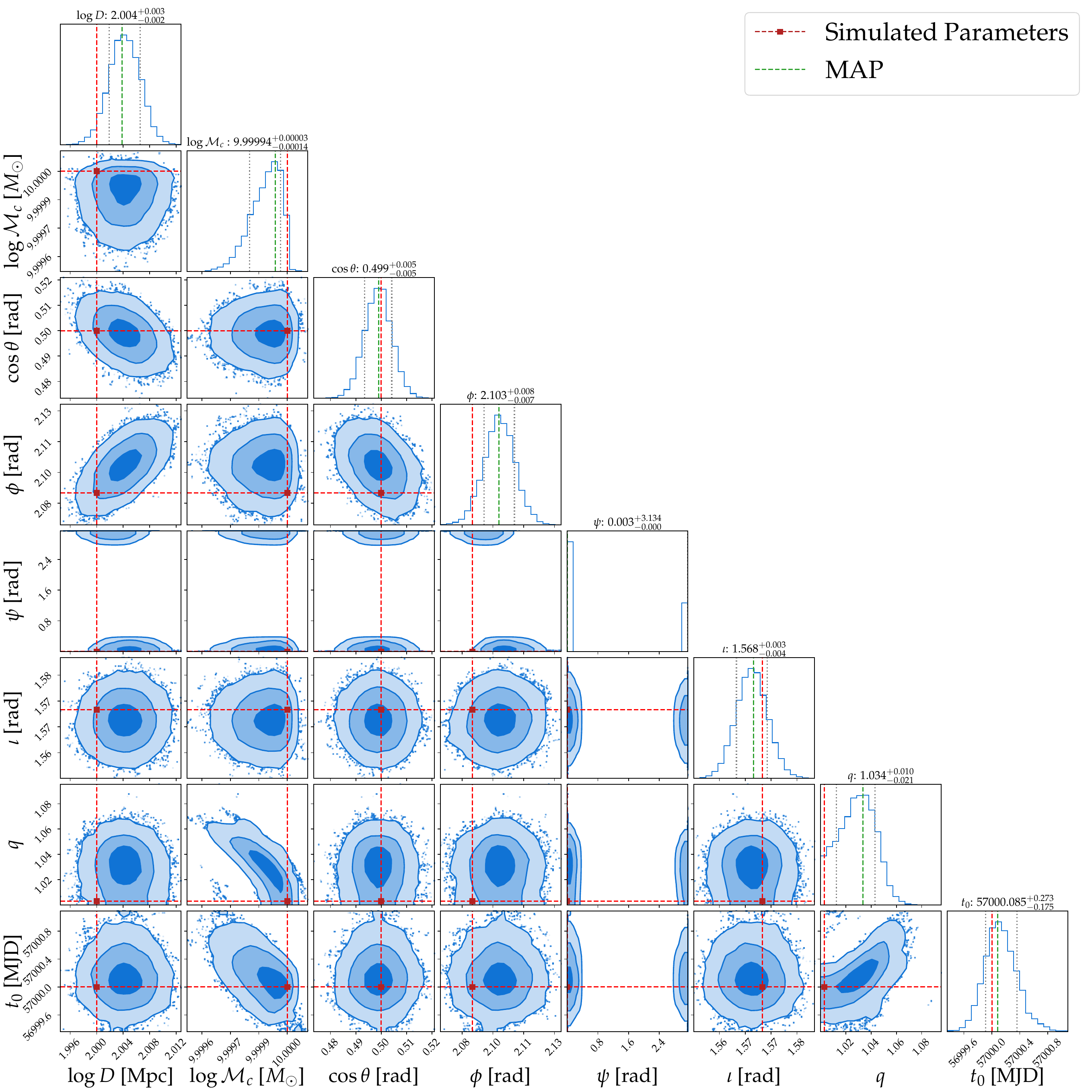}  
  \caption{\justifying
 Posterior distributions of SMBHB merger with null memory parameters from simulation studies on 25 pulsars. Parameters of the binary used for simulation are $\mathcal{M}_c = 10^{10} M_\odot$ and $D_L = 100$ Mpc. Posteriors shown in blue is the recovery of the signal from the \texttt{wn+mem+rn} dataset. Red lines are the simulated values of the SMBHB merger and green lines are the maximum-\textit{a-posteriori} (MAP) estimates from the recovered posterior.
  }
  \label{fig:sim_wn_rn}
\end{figure*}

\begin{figure*}[htbp]
  \centering
  \includegraphics[width=0.97\textwidth]{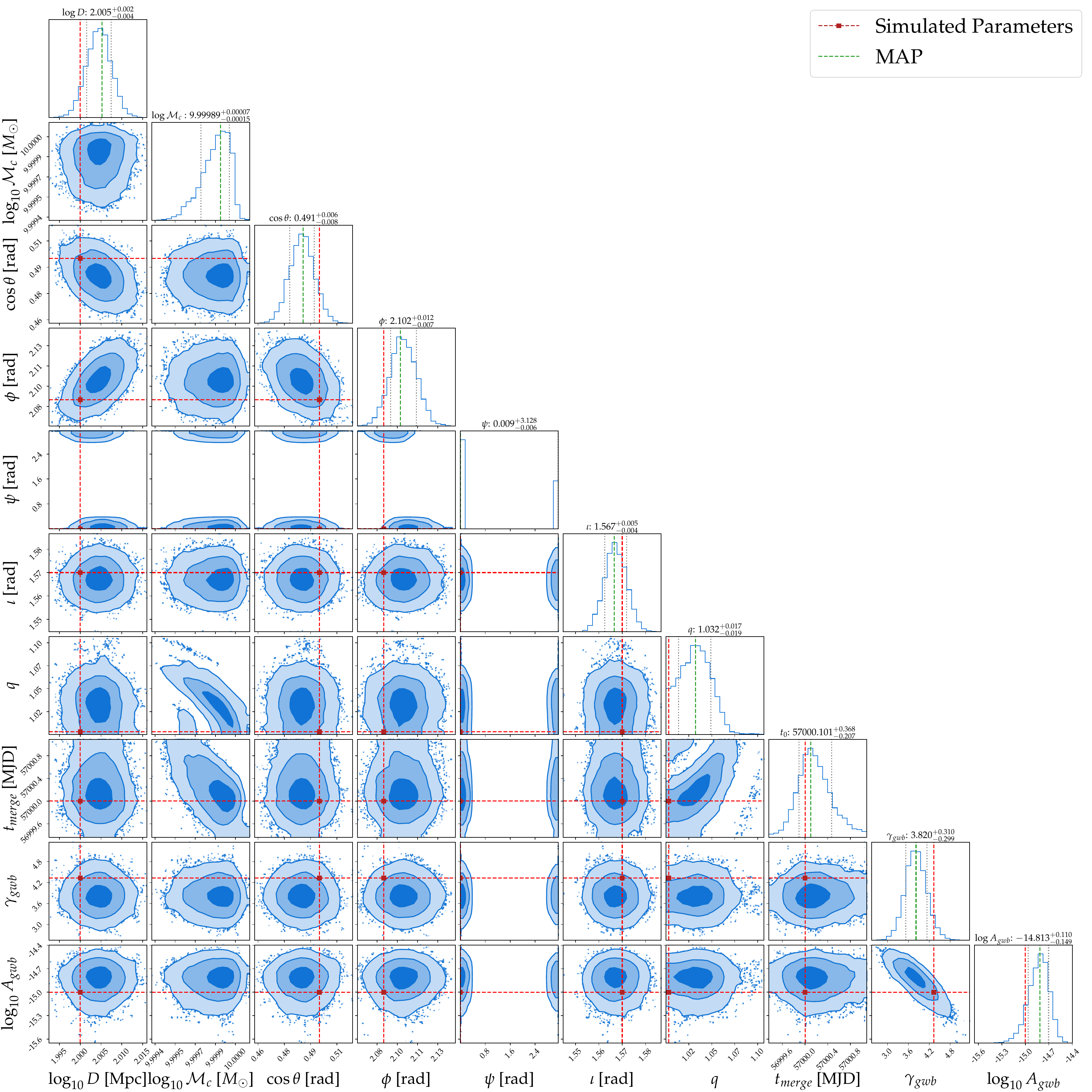}  
  \caption{\justifying
 Posterior distribution of the SMBHB merger from simulation studies on 25 pulsars. Parameters of the binary used for simulation are $\mathcal{M}_c = 10^{10} M_\odot$ and $D_L = 100$ Mpc. The dataset used is \texttt{wn+mem+rn+gwb}. Overplotted red lines indicate the simulated values used in the simulated data, and green lines mark the MAP estimates from the recovered posterior. 
  }
  \label{fig:sim_wn_rn_gwb}
\end{figure*}

\end{document}